\documentclass[fleqn,usenatbib,useAMS]{mnras}

\usepackage[T1]{fontenc}
\usepackage{ae,aecompl}

\usepackage{graphicx}	
\usepackage{amsmath}	
\usepackage{amssymb}	

\newcommand{\bz}{$\langle B_z \rangle$}

\title[Anomalous element distribution in HD\,54879]{
Detection of anomalous element distribution in the extremely slowly 
rotating magnetic O9.7\,V star HD\,54879}

\author[J\"arvinen et al.\ 2021]{
S.~P.~J\"arvinen$^{1}$\thanks{E-mail: sjarvinen@aip.de}
S.~Hubrig$^{1}$,
M.~Sch\"oller$^{2}$,
A.~Cikota$^{3}$,
I.~Ilyin$^{1}$,
\newauthor{
C.~A.~Hummel$^{2}$,
and M. K\"uker$^{1}$}
\\
$^{1}$Leibniz-Institut f\"ur Astrophysik Potsdam (AIP), 
An der Sternwarte~16, 14482~Potsdam, Germany\\
$^{2}$European Southern Observatory, 
Karl-Schwarzschild-Str.~2, 85748 Garching, Germany\\
$^{3}$European Southern Observatory, 
Alonso de C\'ordova 3107, Vitacura, Santiago, Chile
}
\date{Accepted XXX. Received YYY; in original form ZZZ}

\pubyear{2021}

\begin{document}
\label{firstpage}
\pagerange{\pageref{firstpage}--\pageref{lastpage}}
\maketitle

\begin{abstract}

The O9.7\,V star HD\,54879 is currently the only massive magnetic star whose 
magnetic field geometry and rotation period are not constrained. Over the 
last three years, we gathered additional observations of this star, obtained 
using various instruments at several astronomical facilities with, the aim to 
constrain the rotation period and the magnetic field geometry. The new data 
include the first full Stokes vector observations with the PEPSI 
spectropolarimeter, installed at the Large Binocular Telescope. The acquired 
spectropolarimetric observations show a very slow magnetic field variability 
related to the extremely slow rotation of HD\,54879, which is also indicated 
in a dynamical spectrum, displaying variability of the H$\alpha$ line. The 
most intriguing result of our study is the discovery of differences in 
longitudinal  magnetic field strengths measured using different LSD masks 
containing lines belonging to different elements. It is the first time that 
such a differential analysis of the field strength in dependence of the used 
lines is carried out for a magnetic O-type star. Since the LSD Stokes~$I$ 
profiles of the studied O, Si, and He line masks remain stable over all 
observing epochs, we conclude that the detection of different field 
strengths using lines belonging to these elements is related to the 
different formation depths, with the He lines formed much higher in the 
stellar atmosphere compared to the silicon and the oxygen lines, and NLTE 
effects. Our numerical magnetospherical model suggests the presence of 
enhanced gas density that fills the  volume inside the field lines close 
to the star.

\end{abstract}

\begin{keywords}
  stars: individual: HD\,54879 --
  stars: early-type --
  stars: atmospheres --
  stars: variables: general --
  stars: magnetic fields
\end{keywords}



\section{Introduction}
\label{sec:intro}

The origin of the magnetic fields in massive stars is an important problem 
that is not solved yet. A strong support in favour of a scenario that 
magnetic fields in massive stars may be generated by strong  binary 
interaction, i.e.\ in stellar mergers, or during a mass transfer or common 
envelope evolution, was presented in the recent work of 
\citet{frost},
who studied the properties of the massive double-lined system containing the 
magnetic Of?p star HD\,148937. Using multiwavelength observations, they
reported about the age discrepancies of the binary components in the system 
HD\,148937, consistent with the primary component being a rejuvenated merger 
product. The spectral Of?p classification is described by 
\citet{Walborn1972}. 
Five such stars were identified in our Galaxy and magnetic fields were 
detected in all of them.

To answer the principal question on the possible origin of magnetic fields 
in massive stars, it is important not only to build trustworthy statistics 
on the occurrence of magnetic fields, but also on their topology at 
different evolutionary stages. During the last two decades, a considerable 
number of O-type stars have been investigated for magnetic fields, and as a 
result, about a dozen magnetic O-type stars are presently known 
\citep[e.g.][]{Grunhut2017,Schoeller2017}.
The most recent detection of a magnetic field in an O-type star dates 
already seven years back: the presence of a magnetic field in the O9.7V star 
HD\,54879 was reported by 
\citet{Castro2015},
using a combination of FOcal Reducer low dispersion Spectrograph
\citep[FORS\,2;][]{Appenzeller1998}
and high-resolution HARPS\-pol 
\citep[the High Accuracy Radial velocity Planet Searcher in polarimetric 
mode;][]{Snik2008} 
observations. For this 4\,Myr old star, the authors detected a 
$-600$\,G longitudinal magnetic field with a lower limit of the dipole 
strength of $\sim2$\,kG. Despite of several subsequent studies 
\citep[e.g.][and references therein]{HD54_2020}, 
HD\,54879 currently remains the only massive magnetic star whose magnetic 
field geometry and rotation period are not constrained. The reason for 
this is related to the extremely slow rotation of this star: based on the 
variability of the H$\alpha$ line intensity and the longitudinal magnetic 
field,
\citet{HD54_2020} 
concluded that the rotation period is probably longer than 9\,yr. Although 
magnetic O-type stars are known as a class of slowly rotating magnetic 
massive stars, the extremely slowly rotation of HD\,54879 with the second 
longest rotation period after the Of?p star HD\,108 with a rotation period 
of about 50--60\,yr 
\citep{Naze2001},
makes it an important target for studies of the impact of the magnetic 
field on the physical processes taking place in stellar atmospheres and 
inside of stars.

New spectroscopic and spectropolarimetric material has been acquired over the 
last years using various instruments belonging to different astronomical 
facilities. In this work, they are used to study the properties of HD\,54879 
and to put constraints onto the rotation period, the field strength, and the 
field geometry. In this article, we describe the observations in 
Section~\ref{sec:obs} and in Section~\ref{sec:Bz} we present the results of 
the magnetic field measurements and the analysis techniques, and put 
constrains onto the rotation period and the most probable configuration of 
the magnetic field geometry. In Section~\ref{sec:elem}, we report on the 
field strength differences detected using for the measurements lines 
belonging to different elements, followed by a discussion of the current 
status of the numerical simulations, the binarity status, and short-term 
variability in Sections~\ref{sec:MHD}, \ref{sec:bin}, and \ref{sec:pul}. 
Finally, we discuss the obtained results in Section~\ref{sec:dis}.


\section{Observations}
\label{sec:obs}

The observations used in our analysis of the properties of HD\,54879 were 
obtained with several European Southern Observatory (ESO) instruments 
installed at the Very Large Telescope (VLT) on Cerro Paranal, Chile:
the FOcal Reducer low dispersion Spectrograph in spectropolarimetric mode 
\citep[FORS\,2;][]{Appenzeller1998},
the Ultraviolet and Visual Echelle Spectrograph 
\citep[UVES;][]{uves},
and the beam combining instrument Precision Integrated-Optics Near-infrared 
Imaging ExpeRiment
\citep[PIONIER;][]{LeBouquin2011}
on the Very Large Telescope Interferometer 
\citep[VLTI; e.g.][]{Schoeller2007}.
Further, we used the Potsdam Echelle Polarimetric and Spectroscopic Instrument 
\citep[PEPSI;][]{PEPSI}
at the 2$\times$8.4\,m Large Binocular Telescope (LBT) in Arizona to record 
high-resolution spectra in circular and linear polarized light. 
Additionally, we downloaded two archive spectropolarimetric observations 
recorded with the Echelle  SpectroPolarimetric Device for the Observation of 
Stars 
\citep[ESPaDOnS;][]{Donati2006}
installed at the Canada-France-Hawaii Telescope (CFHT). In the following 
subsections, we describe the data acquired with these instruments. 
Finally, we have used the FORS\,2, HARPS\-pol, and ESPaDOnS data already 
presented by 
\citet{HD54_2020}.

\subsection{FORS\,2}

Two new spectropolarimetric observations of HD\,54879  with FORS\,2 were 
obtained on 2020 September~20 and November~15 in the framework of the 
programme 0106.D-0250(A) executed in service mode at UT1 of the VLT.
Similar to observations presented previously by 
\citet{HD54_2020}, 
both FORS\,2 observations were recorded with the GRISM 600B and the 
narrowest available slit width of 0$\farcs$4 to obtain a spectral resolving 
power of $R\approx2,000$ within the observed spectral range from 3250 to 
6215\,\AA. Further, a non-standard readout mode with low gain 
(200kHz,1$\times$1,low) was used, providing a broader dynamic range and hence 
allowing us to reach a higher signal-to-noise ratio ($S/N$) in the individual 
spectra. The position angle of the retarder waveplate was changed from 
$+45^{\circ}$ to $-45^{\circ}$ and vice versa every second exposure. This is 
done to minimize the cross-talk effect and to cancel errors from different 
transmission properties of the two polarized beams. Moreover, the reversal 
of the quarter-wave plate compensates for fixed errors in the relative 
wavelength calibrations of the two polarized spectra. The wavelength 
calibration was carried out using He-Ne-Ar arc lamp exposures. The ordinary 
and extraordinary beams were extracted using standard {\sc iraf} procedures, 
as described by 
\citet{Cikota}.

\subsection{UVES}

Two spectra were recorded on 2020 October~11 and November~25 with UVES 
mounted on UT2 of the VLT within the framework of the programme 
0106.D-0250(B). The obtained spectra have a resolving power $R\approx80,000$ 
in the blue arm and $R\approx110,000$ in the red arm. The spectra cover the 
wavelength ranges 3756--4982\,\AA{} and 5690--9459\,\AA{} and were reduced 
by the ESO Phase~3 UVES 
pipeline\footnote{http://www.eso.org/rm/api/v1/public/releaseDescriptions/163}.

\subsection{Interferometric observations with PIONIER}

HD\,54879 was observed with the beam combining instrument PIONIER on 2018 
November~30 in the framework of the programme 0102.D-0234(C). PIONIER 
combines four beams in the H-band, either from the 8.2\,m Unit Telescopes 
(UTs), or, as in this case, from the 1.8\,m Auxiliary Telescopes (ATs),
which leads to visibilities on six different baselines, as well as four 
closure phase measurements, simultaneously. PIONIER was used without a 
dispersive element, to enhance the sensitivity for this rather faint target.
For brighter objects, the instrument can also operate in a low resolution 
spectroscopic mode to carry out measurements at five different wavelengths 
within the H-band, improving the UV-coverage. The observations of HD\,54879 
were carried out on the AT configuration A0-G1-J2-K0.

\subsection{PEPSI}

We used the Potsdam Echelle Polarimetric and Spectroscopic Instrument 
(PEPSI) at the 2$\times$8.4\,m Large Binocular Telescope (LBT) in Arizona to 
obtain circular and linear polarized spectra of HD\,54879. The spectrograph 
and its spectropolarimetric capabilities have been described in detail by 
\citet{PEPSI}.
The polarimetric mode has a spectral resolution of $R \approx 130,000$. The 
observations were obtained on 2020 December~6 and cover two wavelength 
regions: the crossdisperser II covers the wavelength range 4236--4770\,\AA{} 
and the crossdisperser IV 5391--6288\,\AA. Retarder angles of 45\degr\ and 
135\degr\ were set to record the Stokes~$V$ spectra, whereas the Foster 
prism position angles 0\degr and 90\degr with respect to the North were used 
to obtain Stokes~$Q$, then 45\degr and 135\degr to obtain Stokes~$U$. The 
exposure time for each subexposure accounted for 5~min.

The data reduction was done using the software package {\sc sds4pepsi} 
(``Spectroscopic Data Systems for PEPSI'') based on 
\citet{4A}, 
and described in detail by \citet{SDS4PEPSI}.
It includes bias subtraction and variance estimation of the source images, 
master flat field correction for the CCD spatial noise, scattered light 
subtraction, definition of \'echelle orders, wavelength solution for the ThAr 
images, optimal extraction of image slices and cosmic spike elimination, 
normalisation to the master flat field spectrum to remove CCD fringes and the 
blaze function, a global 2D fit to the continuum, and the rectification of all 
spectral orders into a 1D spectrum.

\subsection{ESPaDOnS}

Recently, two ESPaDOnS polarimetric spectra with a spectral resolution of 
$R\approx65,000$ obtained on 2019 March~15 and September~17 became publicly 
available. We downloaded these already reduced spectra from the CFHT archive.


\section{Constraining the rotation period of HD\,54879 and its
magnetic field geometry}
\label{sec:Bz}

According to previous studies, HD\,54879 is expected to have a rotation 
period of the order of several years
\citep{Shenar2017,Hubrig2019,HD54_2020}.
To determine the periodicities in magnetic massive O-type stars, not only 
long-term monitoring of the variability of the longitudinal magnetic field, 
but also of the variability of equivalent widths of the H$\alpha$ and 
\ion{He}{ii}~4686 lines, which are dominated by stellar wind material in the 
stellar magnetosphere, is usually used 
\citep[e.g.][]{Stahl}.

\subsection{Searching for periodicity using measurements of the longitudinal 
magnetic field}
\label{subsec:pol}

The procedure for the measurement of longitudinal magnetic fields using 
low-resolution FORS\,1/2 spectropolarimetric observations is described
e.g.\ in
\citet[][and references therein]{Hubrig2004a, Hubrig2004b}. 
The results of our measurements are listed in Table~\ref{tab:FORS}.
To obtain robust estimates of the standard errors, we also carried out Monte 
Carlo bootstrapping tests 
\citep[e.g.][]{Steffen2014}.

\begin{table}
\centering
\caption{Longitudinal magnetic field measurements of HD\,54879 using low 
resolution FORS\,2 spectropolarimetric observations. 
The first two columns give the modified Julian dates (MJDs) for the middle 
of the exposure and the signal-to-noise ($S/N$) values of the spectra. 
Magnetic field measurements, those for the entire spectrum and those using 
only the hydrogen lines, are presented in Columns~3 and 4, followed by the 
measurements using all lines in the null spectra, which are obtained from 
pairwise differences from all available Stokes~$V$ profiles so that the real 
polarization signal should cancel out.
The $S/N$ is measured at 4800\,\AA{}.
}
\label{tab:FORS}
\begin{tabular}{ccr@{$\pm$}lr@{$\pm$}lr@{$\pm$}lc}
\hline
\multicolumn{1}{c}{MJD} &
\multicolumn{1}{c}{$S/N$} &
\multicolumn{2}{c}{$\left<B_{\rm z}\right>_{\rm all}$} &
\multicolumn{2}{c}{$\left<B_{\rm z}\right>_{\rm hyd}$} &
\multicolumn{2}{c}{$\left<B_{\rm z}\right>_{\rm N}$} \\
& 
&
\multicolumn{2}{c}{(G)} &
\multicolumn{2}{c}{(G)} &
\multicolumn{2}{c}{(G)}
\\
\hline
59122.37    & 2205 & $-$275 & 95 & $-$424 & 144 &$-$ 51 & 87  \\
58169.26    & 2130 & $-$291 & 103 &$-$649 & 147 & 68 & 106  \\
\hline
\end{tabular}
\end{table}

To measure the longitudinal magnetic field \bz{} from the high-resolution 
spectropolarimetric observations, we employed, similarly to the work of
\citet{HD54_2020},
the least-squares deconvolution (LSD) technique, allowing us to achieve a 
much higher $S/N$ in the polarimetric spectra. For the details of this 
technique and of the calculation of the Stokes~$I$ and Stokes~$V$ parameters,
we refer to the work of
\citet{Donati1997}. 
The line masks were constructed using the Vienna Atomic Line Database 
\citep[VALD3;][]{kupka2011} 
and the stellar parameters $T_{\rm eff}=30.5$\,kK and $\log\,g=4.0$ reported 
by 
\citet{Shenar2017}.
Given the shorter wavelength coverage of the PEPSI spectra compared to the 
spectral coverage of the HARPS and ESPaDOnS spectra, we had to modify our 
line masks by removing the spectral lines not covered by PEPSI. Consequently, 
we re-measured the field values also in the polarimetric spectra from the 
previous analysis by 
\citet{HD54_2020}.

The measurements using the line mask optimised for the PEPSI spectral 
wavelength coverage containing metal and He lines and those using the line 
mask containing exclusively metal lines are listed in the first four columns 
in Table~\ref{tab:hr_Bz} along with the spectropolarimeter identification and 
the modified Julian dates for the middle of each exposure. In all cases, the 
false alarm probability is less than $10^{-6}$. According to 
\citet{Donati1992}, 
a Zeeman profile with false alarm probability (FAP) $\leq 10^{-5}$ is 
considered as a definite detection, $10^{-5} <$ FAP $\leq 10^{-3}$ as a 
marginal detection, and FAP $> 10^{-3}$ as a non-detection. In this table, 
we also indicate the number of lines in each line mask, and the mean effective 
Land\'e factors and the mean wavelengths used for the normalisation for the 
line mask. The calculated LSD Stokes~$I$ and Stokes~$V$ profiles for each 
observing epoch are presented in 
Figs.~\ref{afig:LSDall}--\ref{afig:LSDHe1} in the Appendix.

\begin{table*}
\centering 
\caption{Longitudinal magnetic field \bz{} measured using high-resolution 
HARPS, ESPaDOnS, and PEPSI polarimetric spectra. For each line list the 
number of lines, the mean effective Land\'e factors and the mean 
wavelengths, used for the normalisation for the given line mask, are given.
}
\label{tab:hr_Bz}
\begin{tabular}{cccr@{$\pm$}lr@{$\pm$}lr@{$\pm$}lr@{$\pm$}lr@{$\pm$}l}
\hline 
\multicolumn{1}{c}{Instrument} &
\multicolumn{1}{c}{MJD} &
\multicolumn{1}{c}{} &
\multicolumn{2}{c}{$\left<B_{\rm z}\right>_{\rm all}$} &
\multicolumn{2}{c}{$\left<B_{\rm z}\right>_{\rm met}$} &
\multicolumn{2}{c}{$\left<B_{\rm z}\right>_{\rm \ion{Si}{iii}}$} &
\multicolumn{2}{c}{$\left<B_{\rm z}\right>_{\rm \ion{O}{ii}}$} &
\multicolumn{2}{c}{$\left<B_{\rm z}\right>_{\rm \ion{He}{i}}$} \\
& & &
\multicolumn{2}{c}{(G)} &
\multicolumn{2}{c}{(G)} &
\multicolumn{2}{c}{(G)} &
\multicolumn{2}{c}{(G)} &
\multicolumn{2}{c}{(G)} \\
  \hline 
& & \# of lines &
\multicolumn{2}{c}{70} &
\multicolumn{2}{c}{66} &
\multicolumn{2}{c}{6} &
\multicolumn{2}{c}{15} &
\multicolumn{2}{c}{4} \\
& & $\left<g_{\rm eff}\right>$ &
\multicolumn{2}{c}{1.12} &
\multicolumn{2}{c}{1.11} &
\multicolumn{2}{c}{1.45} &
\multicolumn{2}{c}{1.16} &
\multicolumn{2}{c}{1.20} \\
& & $\left<\lambda\right>$&
\multicolumn{2}{c}{4786\,\AA{}} &
\multicolumn{2}{c}{4732\,\AA{}} &
\multicolumn{2}{c}{4937\,\AA{}} &
\multicolumn{2}{c}{4552\,\AA{}} &
\multicolumn{2}{c}{5230\,\AA{}} \\

  \hline 
  HARPS    & 56770.03 & & $-$388 &  15 & $-$569   &  28 & $-$488 & 14 & $-$539 & 28 & $-$199 & 10 \\
  HARPS    & 57092.09 & & $-$283 &  12 & $-$417   &  23 & $-$333 & 17 & $-$360 & 26 & $-$183 &  9 \\
  HARPS    & 57095.03 & & $-$320 &  17 & $-$430   &  32 & $-$359 & 24 & $-$474 & 26 & $-$245 & 12 \\
  ESPaDOnS & 56970.54 & & $-$312 &  15 & $-$439   &  23 & $-$348 & 13 & $-$425 & 25 & $-$189 & 12 \\
  ESPaDOnS & 56970.58 & & $-$324 &  16 & $-$437   &  33 & $-$319 & 13 & $-$488 & 24 & $-$229 & 11 \\
  ESPaDOnS & 57736.47 & & $-$204 &  14 & $-$250   &  31 & $-$269 & 19 & $-$242 & 21 & $-$181 & 16 \\
  ESPaDOnS & 57758.36 & & $-$255 &  13 & $-$334   &  20 & $-$292 & 14 & $-$248 & 16 & $-$212 & 13 \\
  ESPaDOnS & 57775.46 & & $-$285 &  12 & $-$446   &  22 & $-$285 & 16 & $-$416 & 24 & $-$173 & 10 \\
  ESPaDOnS & 57880.22 & & $-$167 &  12 & $-$229   &  18 & $-$161 & 18 & $-$207 & 29 & $-$160 & 10 \\
  ESPaDOnS & 58007.61 & & $-$122 &   9 & $-$181   &  16 & $-$114 & 17 & $-$120 & 14 & $-$88  &  7 \\
  ESPaDOnS & 58065.51 & &  $-$78 &   7 & $-$88    &  15 & $-$97  &  8 & $-$130 & 17 & $-$97  &  6 \\
  ESPaDOnS & 58128.40 & &  $-$56 &   7 & $-$105   &  14 & $-$130 & 19 & $-$114 & 17 & $-$60  &  9 \\
  ESPaDOnS & 58557.88 & &     45 &   8 &    50    &  14 &    43  & 18 &    50  & 13 &    23  &  7 \\
  ESPaDOnS & 58744.14 & &   $-$5 &   7 & $-$24    &  16 & $-$17  & 19 & $-$70  & 13 & $-$12  &  6 \\
  PEPSI    & 59189.97 & & $-$298 &  15 & $-$313   &  27 & $-$411 & 11 & $-$467 & 23 & $-$229 & 13 \\
\hline 
\end{tabular}
\end{table*}

\begin{figure}
 \centering 
\includegraphics[width=\columnwidth]{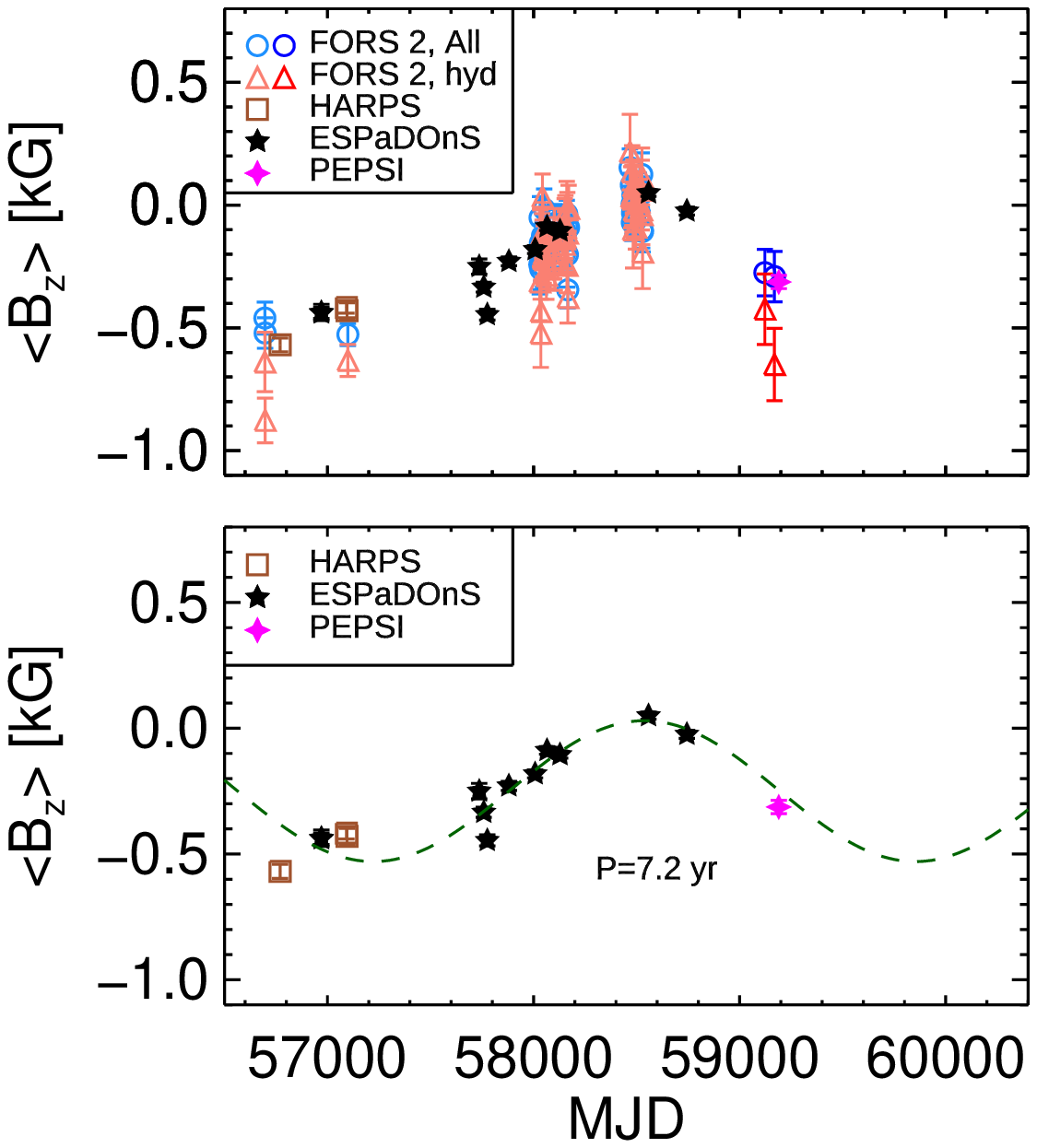}
\caption{
\emph{Upper panel:}
Distribution of the mean longitudinal magnetic field \bz{} values of 
HD\,54879 as a function of MJD between the years 2014 and 2021. The lighter 
colours represent low-resolution data that were published in
\citet{HD54_2020}.
Darker colours are used to present the new measurements. Open circles 
indicate FORS\,2 measurements obtained using the entire spectrum and open 
triangles measurements where only the hydrogen lines were used. All 
high-resolution data have been re-analysed due to different wavelength 
coverage. Open squares and filled stars indicate measurements using HARPS 
and ESPaDOnS high-resolution spectropolarimetric observations and are based 
on the line mask with metal lines. The measurement obtained using the PEPSI 
spectra is indicated by a magenta diamond. 
\emph{Lower panel:}
As above, but showing only the \bz{} obtained from high-resolution observations.
The dashed dark green line shows a cosine curve with the period of 7.2\,yr.
}
   \label{fig:Bevol}
\end{figure}

The distribution of the measured mean longitudinal magnetic field values 
using spectra obtained with different spectropolarimeters as a function of 
MJD is presented in Fig.~\ref{fig:Bevol}. While the mean longitudinal 
magnetic field measured in metal lines using the HARPS spectra in 2014 April 
showed a strength of about $-$569\,G, the field strength values were 
gradually decreasing over the subsequent years and changed their polarity from 
negative to positive around 2019. The strongest longitudinal field with a 
positive polarity of about 50\,G was measured using the line mask with metal 
lines on the night of 2019 March~15. The measurements on 2019 September~17 
showed that the field values reversed the sign again, showing negative 
polarity and reaching a field strength of about $-$300\,G in our last 
observations with PEPSI on 2020 December~6. The distribution of the field 
strengths measured on the low-resolution FORS\,2 polarimetric spectra shows 
a larger dispersion of the data points, but is roughly similar to that of 
the measurements using high-resolution spectropolarimetric observations. We 
observe that FORS\,2 field strengths from using only the hydrogen lines are 
in absolute values higher than those using the entire spectra. This effect is 
already visible in previous studies of a number of magnetic early B-type stars 
\citep[e.g.][]{Hubrig2009,Hubrig2017}.

Generally, magnetic massive O- and B-type stars exhibit a smooth, single-wave 
variation of the longitudinal magnetic field during the stellar rotation 
cycle. The approximately sinusoidal variation of \bz{} and the ratio of the 
values of the \bz{} extrema in previously studied stars suggest that there 
is an important component of the field that is dipolar. Assuming that the 
magnetic field of HD 54879 has a pure dipolar configuration and that the 
negative field extremum is indeed around $-$570\,G and not at a lower value, 
we fitted a cosine curve to the observed distribution of data points 
obtained from the high-resolution spectropolarimetric observations and 
determined a stellar rotation period of 7.2\,yr. Certainly, further 
monitoring of the magnetic field variability is needed to determine the 
rotation period with more confidence. The result of our fitting procedure 
is presented in the lower panel of Fig.~\ref{fig:Bevol} by a dark dashed 
green cosine curve corresponding to the 7.2\,yr rotation period.

This simple cosine  curve presenting the variability of the \bz{} values as 
a function of the rotation phase has an amplitude 
$A_{\left<B_{\rm z}\right>} = 280\pm10$\,G and a mean value 
$\overline{\left<B_{\rm z}\right>} = -250\pm10$\,G. Using the values from 
the fit of the longitudinal magnetic field over the rotation period, one can 
determine a minimum dipolar magnetic field strength $B_d$ of 1950\,G, using 
a limb darkening parameter of 0.2. As we do not know the inclination of the 
star and can not estimate it due to HD\,54879's long rotation period, we can 
not determine the actual dipolar field strength, which e.g.\ would be 
around 6\,kG for an inclination $i$ of 10$^{\circ}$. As already discussed by 
\citet{silva-ngc},
magnetic studies of several O-type stars indicate that only one magnetic 
pole is well visible while the star rotates, implying that the magnetic 
field structure over the fraction of their invisible surface remains 
unconstrained.

\begin{figure}
 \centering 
\includegraphics[width=\columnwidth]{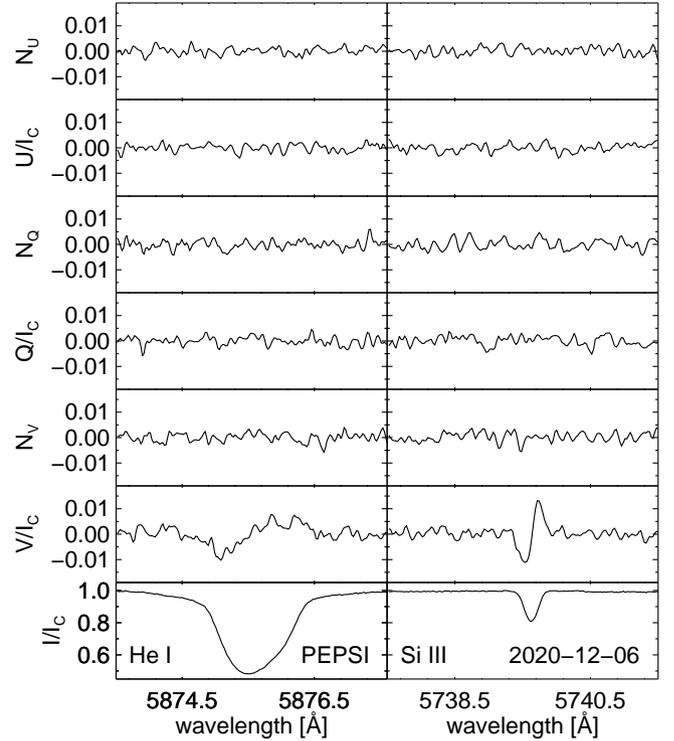}
\caption{
The four Stokes parameters and their null profiles for two individual lines, 
\ion{He}{i} at 5875.6\,\AA{} and \ion{Si}{iii} at 5739.7\,\AA{}, in PEPSI 
observations obtained on 2020 December~6. In the spectra recorded in circular 
polarized light, clear Zeeman signatures are detected for both lines whereas 
the spectra recorded in linear polarized light appear flat.}
   \label{fig:IVQU}
\end{figure}

As the mean longitudinal magnetic field is defined as the average over the 
visible stellar hemisphere of the line-of-sight component of the magnetic 
vector, only the availability of all four Stokes parameters ($I$, $V$, $Q$, 
and $U$) permits a study of the longitudinal and transversal field 
components to determine the real topology of stellar magnetic fields. That 
is, to prove whether massive stars possess dipole-like fields or more complex 
fields with contributions from higher-order harmonics. Measurements of 
linear polarization are also a powerful means to study asymmetries in 
stellar winds. Since HD\,54879 possesses a magnetosphere 
\citep[e.g.][]{HD54_2020},
a study of the variability of the linear polarization in photospheric and 
wind sensitive lines is of great importance to investigate the field geometry 
and circumstellar environment. However, no high-resolution linear 
polarization measurements were carried out for any magnetic O-type star so far.

In Fig.~\ref{fig:IVQU} we present the first PEPSI observations of the full 
Stokes vector for HD\,54879. These observations were carried out with the 
LBT in binocular mode in the night of 2020 December~6. At that epoch, the 
strength of the longitudinal magnetic field was about $-$300\,G. However, 
despite our expectations, no linear polarization signatures were discovered 
in the PEPSI spectra, which were recorded at a $S/N$ of about 500 in the 
corresponding Stokes~$I$ spectra. In this figure we show the PEPSI spectra 
for all four Stokes parameters and their null spectra in the vicinity of two 
spectral lines, the \ion{He}{i} line at 5875.6\,\AA{}, which is sensitive to 
the circumstellar environment, and the photospheric line \ion{Si}{III} at 
5739.7\,\AA{}. The asymmetrical profile of the \ion{He}{i} 5876 triplet 
presented in Fig.~\ref{fig:IVQU} can indicate the presence of an 
asymmetrical wind. A similar asymmetrical line profile is also detected in 
the \ion{He}{i} 7065 triplet. On the other hand, we cannot exclude that due 
to the presence of the magnetosphere around HD\,54879, the red side of the 
line profiles of the \ion{He}{i} 5876 and \ion{He}{i} 7065 lines may be 
partially filled in with emission: these lines appear in emission in the 
spectra of the Of?p star NGC1624-2, which has the strongest magnetic field 
among O type stars, with a dipole strength $\ge16$\,kG 
\citep{silva-ngc},
and their intensity varies in phase with that of the emission line 
\ion{He}{ii} 4686, sensitive to the circumstellar environment.

Unfortunately, the line most dominated by the magnetospheric wind material,
H$\alpha$,  is not covered in the PEPSI spectra presented here. Therefore, 
we plan to record the circumstellar lines H$\alpha$, H$\beta$, and 
\ion{He}{ii} at 4686\,\AA{}, sensitive to the stellar wind material in the 
magnetosphere, during future PEPSI linear polarization observations. To 
explore the structure of the circumstellar environment, such observations 
should be carried out at the epoch of the best visibility of the negative 
magnetic pole.

\subsection{Searching for periodicity using spectroscopic data}
\label{subsec:spectr}

Previous studies of magnetic massive stars indicated that the presence of  
variable emission in H$\alpha$ and the \ion{He}{ii} line at 4686\,\AA{} is 
indicative of circumstellar structures around magnetic O-type stars and 
related to their magnetospheres. Further, a correlation exists between the 
absolute value of the mean longitudinal magnetic field and the strength of 
these emissions in the sense that the strongest emissions appear at phases of 
maximum absolute value of the mean longitudinal magnetic field. While the 
\ion{He}{ii}~$\lambda$4686 line profile in the spectra of HD\,54879 is 
observed exclusively in absorption, the emission line profiles of H$\alpha$ 
are highly variable, changing between a triple-peak and a double-peak 
emission line profile, indicating a composite structure of the surrounding 
circumstellar material 
\citep[e.g.][]{HD54_2020}.

\begin{figure*}
 \centering 
\includegraphics[width=0.3\textwidth]{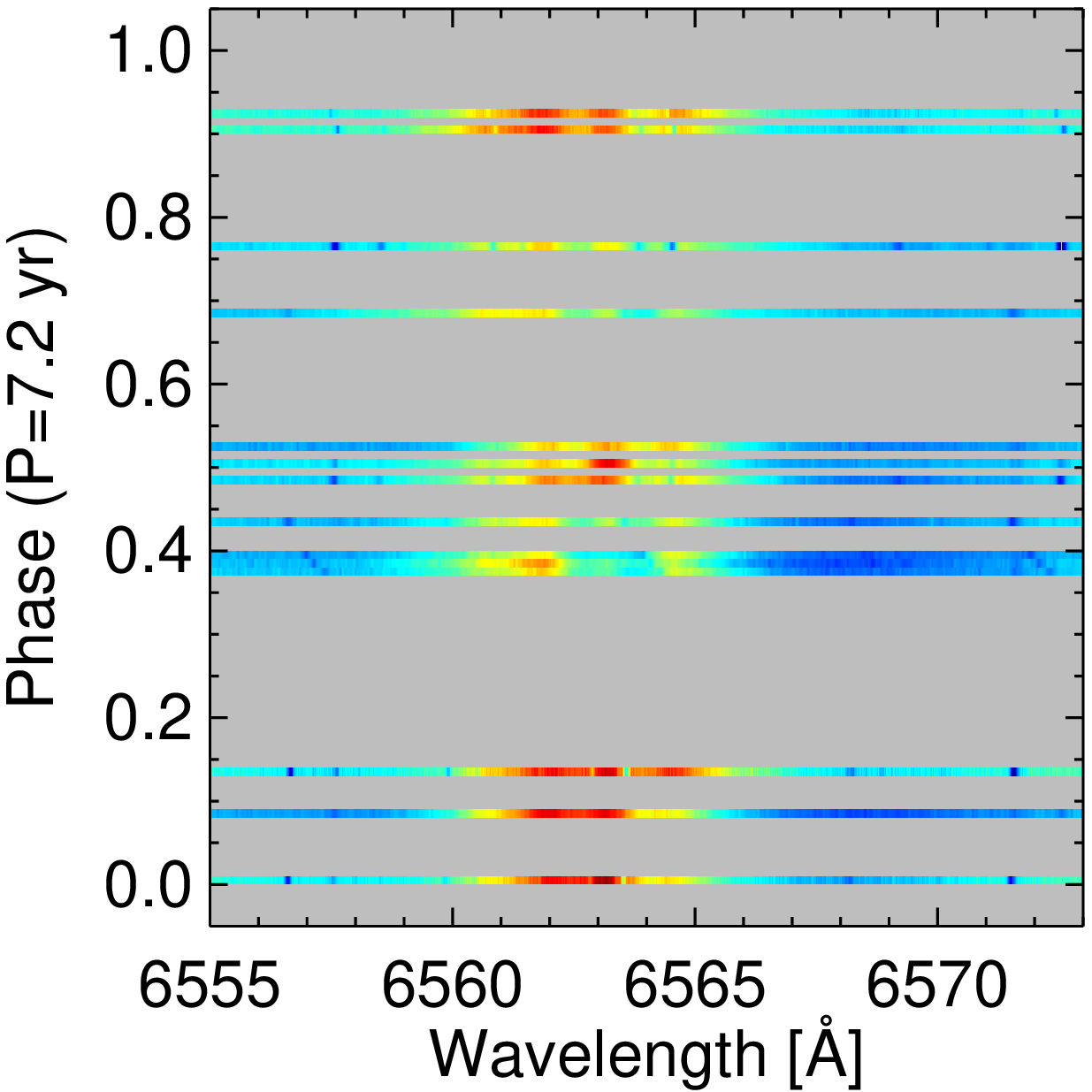}
\includegraphics[width=0.3\textwidth]{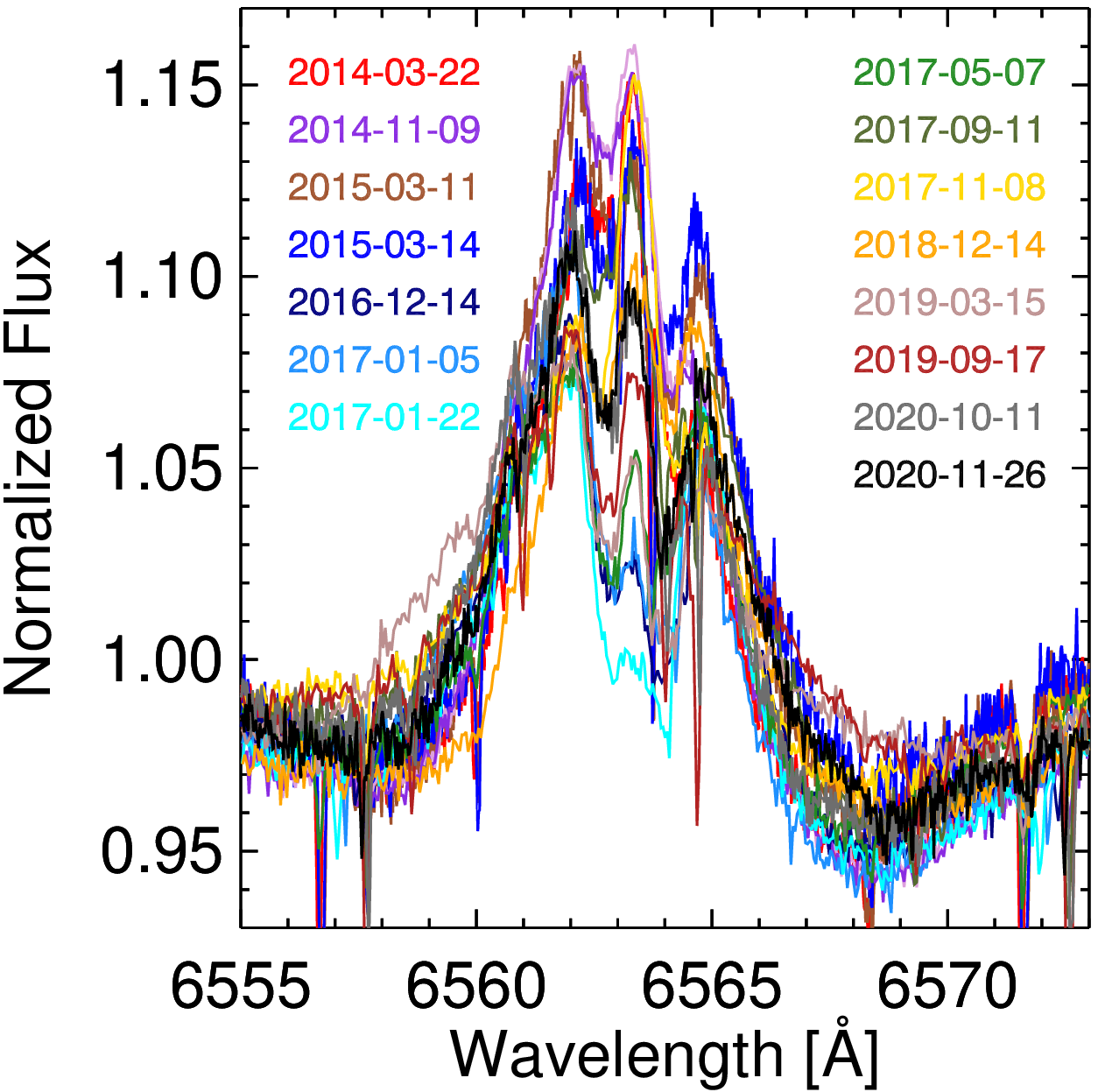}
\includegraphics[width=0.3\textwidth]{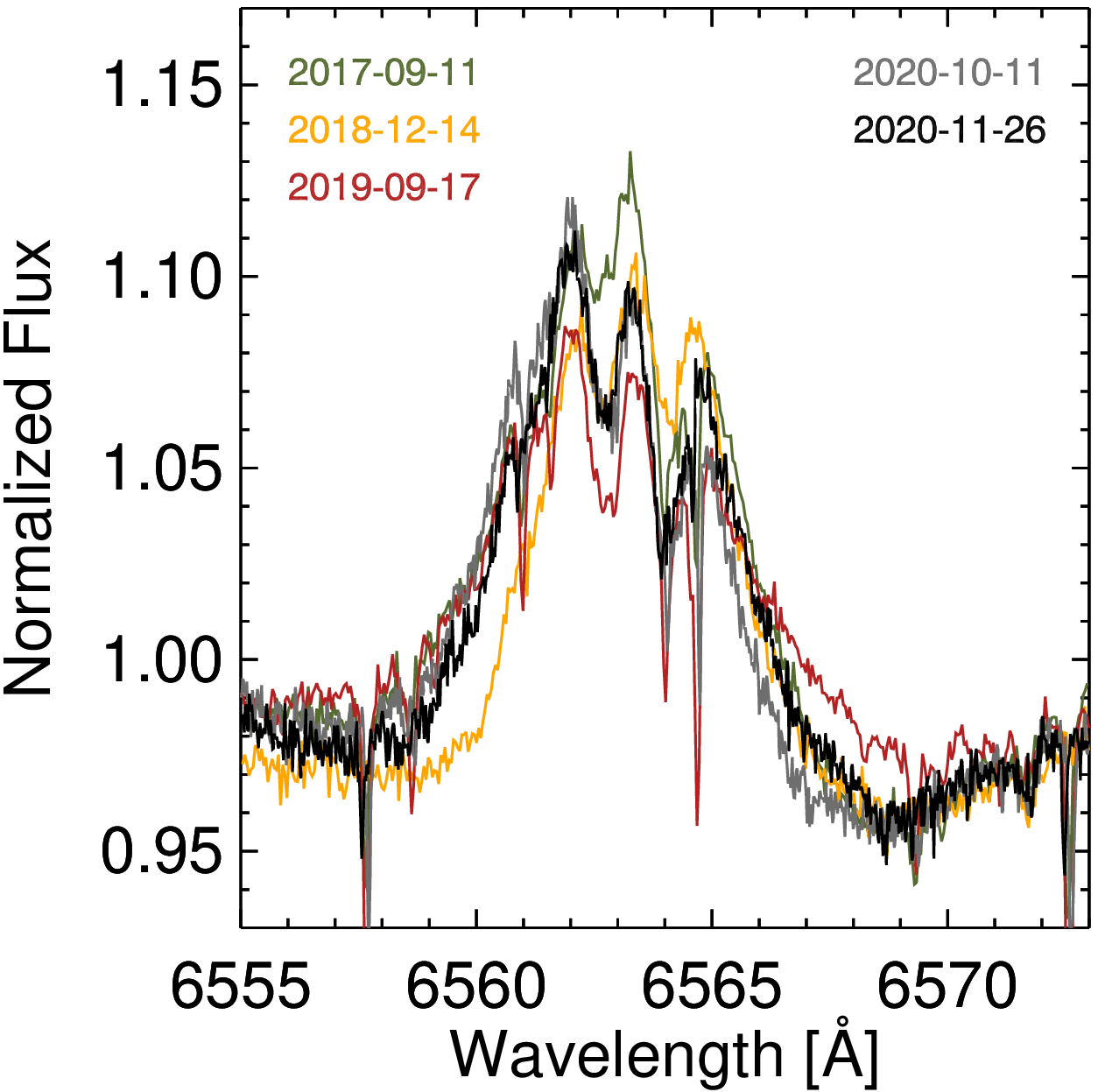}
\caption{
\emph{Left panel:}
A dynamical spectrum showing the variability of the H$\alpha$ profile seen 
in HARPS, ESPaDOnS, and UVES spectra. The observations are phased assuming 
$P=7.2$\,yr. The red colour corresponds to the strongest emission, while the 
blue colour shows the H$\alpha$ profile wings appearing in absorption. The 
strongest emission is detected at the epochs when the magnetic field is 
strongest, but the profiles show also short term variability, already 
mentioned in the previous work by \citet{HD54_2020}.
\emph{Middle panel:}
The variability of the H$\alpha$ profiles in HARPS, EsPaDOnS, and UVES 
spectroscopic observations. 
\emph{Right panel:}
A few ESPaDOnS observations of H$\alpha$ profiles showing similarity with 
the profiles from our recent UVES observations. 
         }
   \label{fig:Halpha}
\end{figure*}

In the left panel of Fig.~\ref{fig:Halpha}, we present all available 
high-resolution spectroscopic observations of the H$\alpha$ line, including 
the two most recent UVES observations, as a dynamical spectrum phased with the 
period of 7.2\,yr. The displayed H$\alpha$ profiles recorded in the UVES 
observations show lower emission levels than the H$\alpha$ profile observed 
in the HARPS spectrum obtained in 2014 (Fig.~\ref{fig:Halpha}, middle panel),
when the maximum longitudinal magnetic field strength was observed. In the 
right panel we show only a few H$\alpha$ line profiles recorded with 
ESPaDOnS resembling the most recent UVES observations. The presented 
differences in the emission levels of the line profiles indicate that we are 
probably a couple of years away from the epoch that allows the best 
visibility of the negative magnetic pole.
 
\begin{figure}
 \centering 
\includegraphics[width=0.35\textwidth]{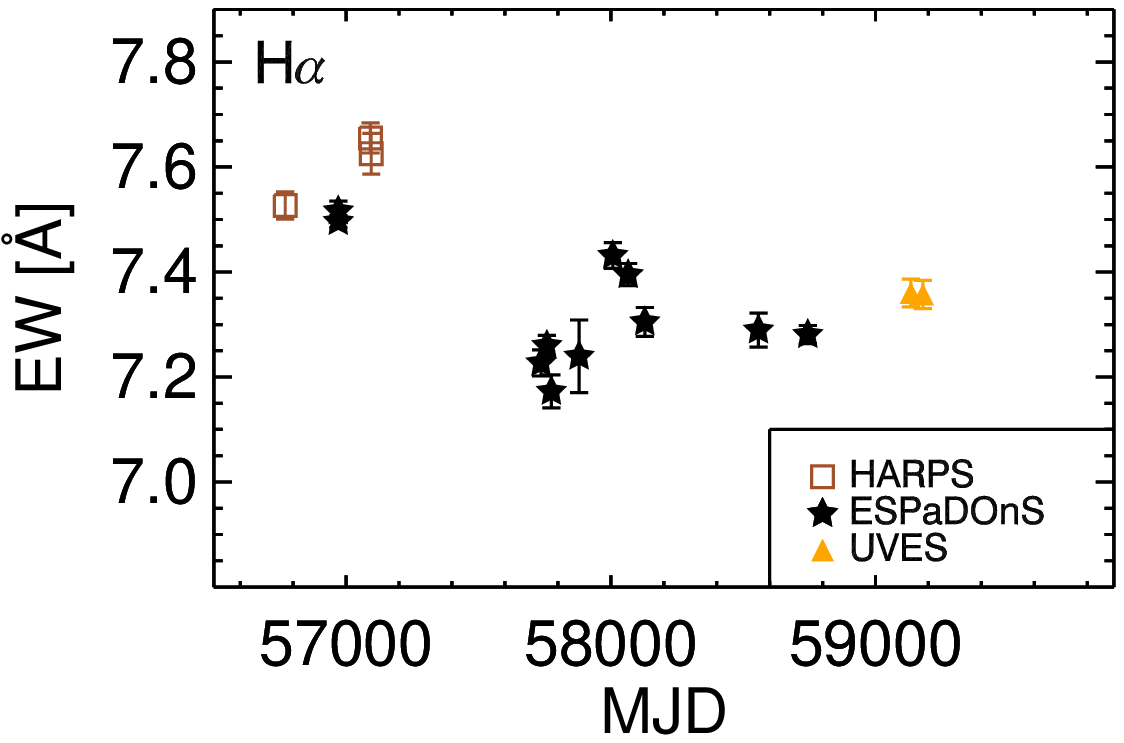}
\includegraphics[width=0.35\textwidth]{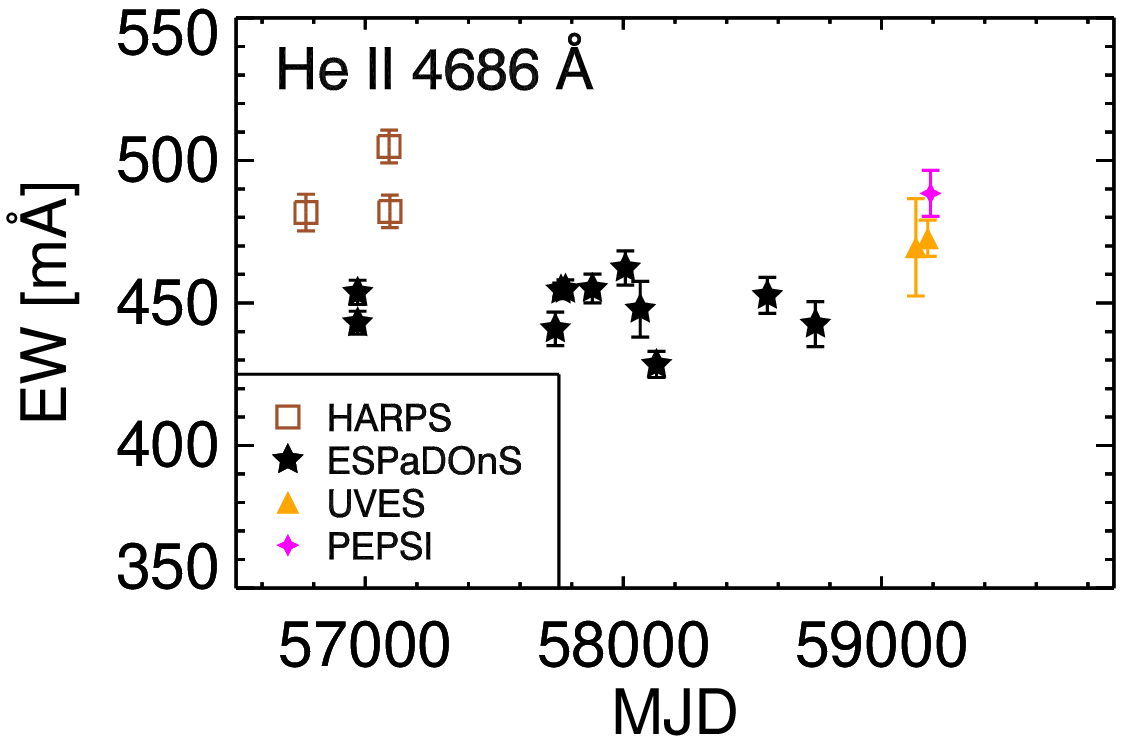}
\caption{
Variability of the H$\alpha$ and \ion{He}{ii}~4686 lines observed on the 
available high-resolution HARPS, ESPaDOnS, UVES, and PEPSI spectra. As the 
H$\alpha$ line exhibits a very complex variable line profile, to calculate 
the EWs, the integration was carried out over the most variable part of the 
line profile between 6559 and 6566\,\AA.
         }
   \label{fig:EWHalpha}
\end{figure}

The variability of the equivalent widths (EWs) of the circumstellar emission 
line H$\alpha$ and the \ion{He}{ii}~$\lambda$4686 line, sensitive to stellar 
wind material, is frequently used to constrain the rotation period in 
magnetic O-type stars 
\citep[e.g.][]{Stahl}.
In Fig.~\ref{fig:EWHalpha} the EWs of these two lines are plotted as a 
function of MJD. As can be seen in this figure, the EWs of the H$\alpha$ and 
\ion{He}{ii}~$\lambda$4686 lines were the strongest at the epoch of the best 
visibility of the negative magnetic pole. The EW values measured using 
ESPaDOnS spectra appear systematically lower compared to those using HARPS 
spectra. This can be due to the presence of scattered light or the lower 
resolution of the ESPaDOnS spectra with $R\approx65,000$, while the spectral 
resolution of the HARPS spectra is about $110,000$.

Considering the distribution of the measurements shown in 
Fig.~\ref{fig:EWHalpha}, it appears that the EWs of both lines show an 
increase during the most recent UVES and PEPSI observations, but further 
monitoring is necessary to confirm this trend. Additionally, we observe a 
rather large scatter in the EW values in the ESPaDOnS spectra during the 
period between MJDs 57740 and 58130, when the longitudinal magnetic field 
was decreasing from $-$250 to $-$105\,G. According to 
\citet{HD54_2020},
the H$\alpha$ lines in the spectra of HD\,54879 show variability on time 
scales of weeks and even hours. It is possible that this variability is 
related to the wind or the immediate environment of the star. The impact of 
the magnetic field of HD\,54879 on its wind characteristics was discussed in 
detail by 
\citet{Hubrig2019},
who used the {\sc nirvana} magnetohydrodynamical (MHD) code 
\citep{nirvana}
to study the distribution of the trapped gas within the magnetosphere and in 
particular the presence of outbursts. Their simulations showed that outside 
of the confinement radius, gas still concentrates in elongated structures
along the magnetic field lines, but is ejected away from the star in 
episodic outbursts.


\section{Field strength differences in measurements from different elements}
\label{sec:elem}

The {\sc nirvana} MHD simulations of the slowly rotating magnetic O-type stars 
NGC\,1624-2 
\citep{silva-ngc}
and HD\,54879 
\citep{HD54_2020}
showed that these stars form dynamical magnetospheres, in which material 
flows along closed magnetic field loops from both magnetic poles, colliding
near the magnetic equator and then falling back on to the stellar surface, 
leading to complex flows in the magnetosphere. The study of the strongly 
magnetic Of?p star NGC\,1624-2 by 
\citet{silva-ngc}
showed opposite variability of the hydrogen and helium emission lines in 
comparison to the variability of the photospheric absorption lines 
\ion{He}{i}~$\lambda$4713 and \ion{C}{iv}~$\lambda$5801. The emission lines 
become stronger when the magnetically confined cooling disc around the 
magnetic equator is seen closer to face-on, while the emission contribution 
is reduced when the cooling disc is seen closer to edge-on. The photospheric 
absorption lines appear stronger when the cooling disc is seen closer to 
edge-on and weaker when the cooling disc is seen closer to face-on. In that 
study, an interesting fact was reported with respect to the LSD magnetic 
field measurements using a line mask with photospheric absorption lines 
including the Si lines, or a line mask without the Si lines. The measured 
magnetic field strength became lower if the Si lines were included in the 
line mask. To explain these measurements, it was suggested that Si is 
inhomogeneously distributed on the stellar surface with a region of Si 
concentration located in the vicinity of the magnetic equator. 
Alternatively, a filling of the Si lines by emission could take place in 
rotation phases where the cooling disc around the magnetic equator is seen 
closer to face-on.

  \begin{figure}
 \centering 
\includegraphics[width=\columnwidth]{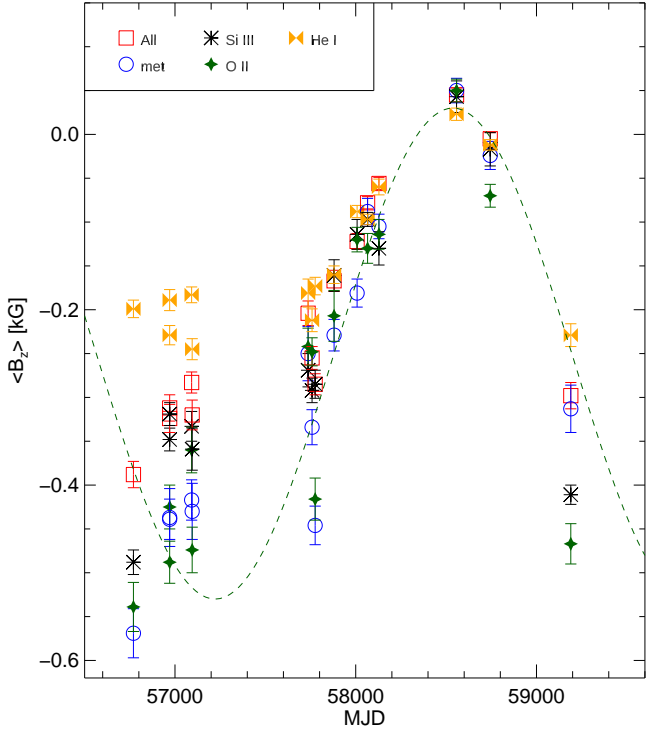}
                \caption{
Distribution of the mean longitudinal magnetic field values \bz{} calculated 
for the different epochs listed in Table~\ref{tab:hr_Bz} using different 
line masks. As in Fig.~\ref{fig:Bevol}, the dashed dark green line shows a 
cosine curve based on measurements using only metal lines with the period of 
7.2\,yr.
         }
   \label{fig:Bzelem}
\end{figure}

\begin{figure*}
\centering 
\includegraphics[width=0.7\textwidth]{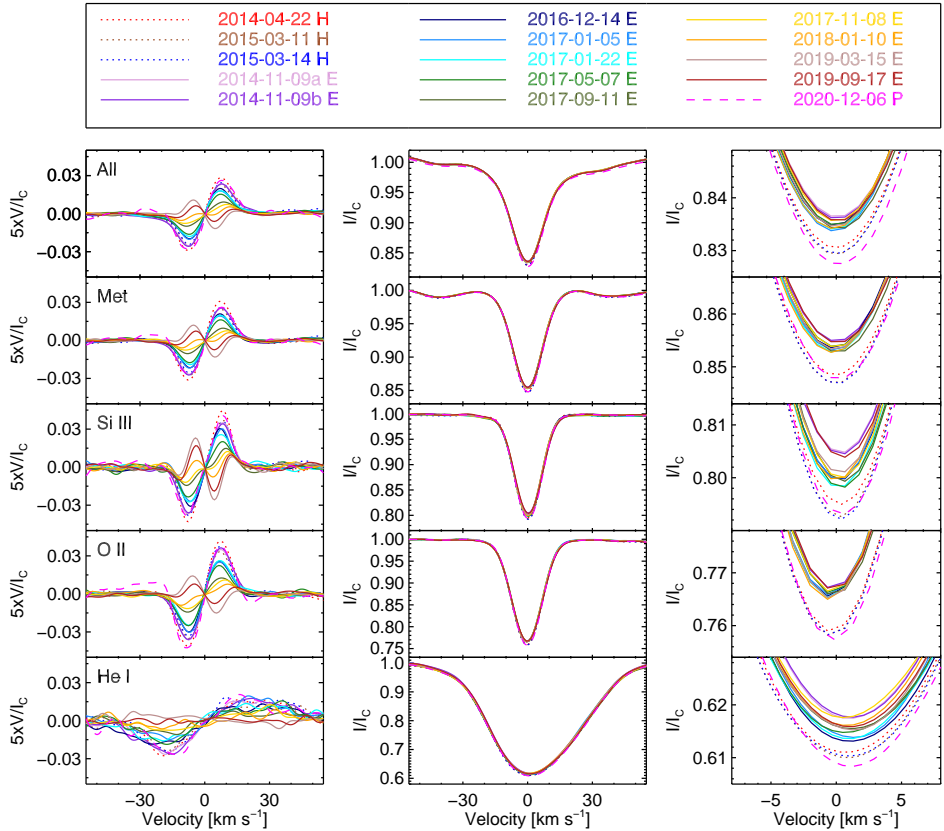}
\caption{
Overplotted LSD Stokes~$V$ profiles (left panel) and LSD Stokes~$I$ profiles 
(middle panel) computed using high-resolution polarimetric spectra recorded 
with HARPS (dotted lines), ESPaDOnS (solid lines) and PEPSI (dashed lines). 
In the right panel we plot the line cores of the Stokes~$I$ profiles at an
enlarged scale for better visibility of the differences between the
observations obtained at different epochs. On the top of this figure, we 
present the dates of the observations and the identification of the lines used.
}
\label{fig:VIevo}
\end{figure*}

\begin{figure}
\centering 
\includegraphics[width=0.45\textwidth]{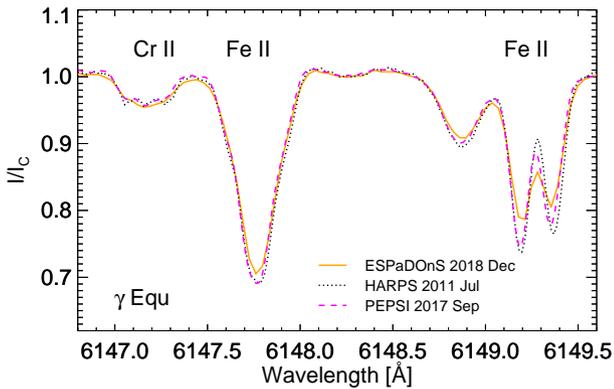}
\caption{
Observations with ESPaDOnS, HARPS, and PEPSI of the spectral region 
containing  the Zeeman doublet \ion{Fe}{ii} at 6149.25\,\AA{} in the spectra 
of $\gamma$\,Equ.
}
\label{fig:scat}
\end{figure}

It is well-known that the majority of magnetic Ap and Bp stars with 
large-scale organised magnetic fields exhibit variations in local abundances 
of various chemical elements, both horizontally and vertically, with a 
patchiness varying from one element to another. Therefore, in these stars, 
the measured magnetic field strengths depend on the distribution of the 
elements whose lines are used for the field measurements. In addition, due 
to the non-uniform horizontal distribution of chemical elements on their 
surfaces, the line profiles of inhomogeneously distributed elements show 
distinct variability over the rotation period and their variability can be 
used to determine the rotation period. Two different mechanisms are usually 
invoked to explain the presence of chemical spots in Ap and Bp stars, 
radiatively driven diffusion 
\citep{Michaud}
and stellar wind fractionation 
\citep{hungergroote}.

To test whether the LSD magnetic field measurements using individual masks 
with lines belonging to different elements show different results for 
HD\,54879, as it was found for NGC\,1624-2 by adding Si lines, we constructed 
three line  masks for the elements Si, O, and He. Since numerous spectral 
lines in the VALD3 database
\citep[][]{kupka2011}
do not have known effective Land\'e factors, a default value g$_{\rm eff}$=1.0 
is frequently used. However, a reasonable comparison between the 
measurements carried out using lines belonging to different elements can 
only be done if their $g_{\rm eff}$ values, i.e.\ their magnetic field 
sensitivities, are known. Among the most clean blend-free spectral lines 
with known effective Land\'e factors, the \ion{O}{ii} lines are most 
numerous, followed by \ion{Si}{iii} and \ion{Si}{iv} lines, and by 
\ion{He}{i} lines. The measurements using these three masks are listed in 
Table~\ref{tab:hr_Bz} in the last three columns and their distribution as a 
function of MJD is presented in Fig.~\ref{fig:Bzelem} along with the 
measurements using line masks for all lines and metal lines, which include 
spectral lines with and without known Land\'e factors.

The observed trends of the longitudinal magnetic field strengths with 
elapsed time obtained using different masks appear qualitatively similar, but 
remarkable differences in the measurements using the individual line masks, in 
particular for the He mask, are well visible on a few epochs. The dispersion 
in field strengths is noticeably large at the epochs close to the best 
visibility of the negative magnetic field pole. The measured field strength 
\bz{} using the He lines is systematically lower than the field strength 
measured using the other masks and the field strength \bz{} using the O 
lines is systematically higher than the field strengths obtained using the 
Si lines. Interestingly, a similar behaviour was recently reported for the 
rapidly oscillating Ap star HD\,166473 
\citep{roAp},
for which a dispersion in field strengths was found to be remarkably large 
in the regions close to the magnetic field poles. However, HD\,166473 is a 
pulsating star showing a strong vertical abundance stratification of iron 
peak and rare earth elements. For such stars, the existence of a relation 
between the magnetic field strength and its orientation and vertical element 
stratification is expected. 

In Fig.~\ref{fig:VIevo} we show in the left and in the middle panel the time 
evolution of the LSD Stokes~$V$ and $I$  profiles calculated for all line 
masks. The LSD Stokes~$V$ profiles change their shape in accordance with the 
measured longitudinal magnetic strengths. The magnetic field has most of the 
time a negative polarity, with the exception of 2019 March~15, when the 
field is clearly positive. We observe crossover profiles on 2019 September~17,
when the magnetic field changes polarity (see also the plots in 
Appendix~A). 
A description of the crossover effect can be found in the work of 
\citet{Mathys1995}.

Despite the clear variability of the LSD Stokes~$V$ profiles, the LSD 
Stokes~$I$ profiles in all line masks constructed for the analysis of 
HD\,54879, remain stable over all observed epochs from 2014 to 2020 (middle 
panel of Fig.~\ref{fig:VIevo}), and only weak profile changes in the line 
cores are observed (right panel of Fig.~\ref{fig:VIevo}). This is in 
contrast to the behaviour of the Stokes~$I$ profiles in Ap and Bp stars.
An inspection of the line cores indicates a clear separation between the 
core depths recorded with ESPaDOnS and those recorded with HARPS and PEPSI. 
The different depths clearly visible in the overplotted line cores in the 
spectra obtained using different spectropolarimeters are probably caused by 
the much lower spectral resolution of ESPaDOnS compared to the spectral 
resolution of HARPS and PEPSI. As we show in Fig.~\ref{fig:scat}, the 
differences in core depths can also be due to the presence of scattered 
light in the ESPaDOnS spectropolarimeter. In this figure we show ESPaDOnS, 
HARPS, and PEPSI observations of the spectral region containing the Zeeman 
doublet \ion{Fe}{ii} at 6149.25\,\AA{} in the spectra of $\gamma$\,Equ. 
Since this star with very sharp lines has an extremely long rotation period 
of at least 95.5\,yr 
\citep{gammaEquper},
no significant changes in the line profiles are expected over a couple of 
years. Again, we observe markedly lower depths of the line profiles recorded 
with ESPaDOnS compared to the line depths recorded with HARPS and PEPSI. 

Because of the different depths between the ESPaDOnS spectra and the higher 
resolution HARPS and PEPSI spectra, the differences in the line cores have 
to be considered separately. A most obvious conclusion with respect to the 
degree of variability follows from the consideration of the Si lines, which 
show the largest depth differences in higher and lower resolution spectra, 
followed by the He lines. The O lines appear to be the most stable lines. 
Further, we observe a trend for lower core depths to appear in observations 
carried out in 2014 and 2019, i.e.\ at epochs when \bz{} is the 
strongest or when \bz{} is around zero.

On the surface of magnetic He-rich early-B type stars analyses of the 
inhomogeneous element distribution usually indicate the presence of He spots 
located close or around the magnetic poles and a high concentration of Si 
spots located close to the magnetic equator 
\citep[e.g.][]{Hubrig2017}.
As the O9.7\,V star HD\,54879 is hotter than the magnetic early-B type stars, 
the stellar wind fractionation scenario predicting an enrichment of helium at 
the poles, which is dependent on the ionization of H and He in the stellar 
wind 
\citep{hungergroote},
is not expected to work in O-type stars usually showing higher 
mass-loss rates 
\citep{vink}.
According to 
\citet{krticka},
early-B type stars have line driven winds with mass-loss rates of the order 
of $10^{-9}M_{\odot}$yr$^{-1}$. This mass loss rate, as suggested by 
\citet{porterskouza},
presents a maximum value to enable the stellar wind fractionation to operate. 
On the other hand, the empirical mass-loss rate estimation in HD\,54879 by 
\citet{Shenar2017}
using the two most prominent wind features, the resonance lines 
\ion{C}{iv}~$\lambda\lambda$1548, 1551, and 
\ion{Si}{iv}~$\lambda\lambda$1394, 1403 recorded with Hubble Space Telescope 
UV observations, showed a much lower mass-loss rate than the theoretical 
estimates for early B-type stars by 
\citet{krticka},
namely $10^{-10.2}M_{\odot}$yr$^{-1}$. 

Still, as we do not detect noteworthy variability in the Stokes~$I$ line 
profiles belonging to O, Si, and He, the stellar wind fractionation scenario 
may not seem to be plausible and another interpretation for the detection 
of different magnetic field strengths measured for different elements needs 
to be found. As mentioned above, previous studies of magnetic O-type stars 
report line profile variability belonging to different elements with the 
rotation period 
\citep[e.g.][]{Stahl,wade2015,silva-ngc}.
This variability is interpreted not in the context of variations of the 
chemical abundance, but as due to variable contamination by circumstellar 
matter. However, the detected stability of the Stokes~$I$ line profiles 
calculated for different line masks in the spectra of HD\,54879 indicates 
that the studied spectral lines are not strongly affected by emission from 
the stellar wind or material trapped in the magnetosphere, suggesting that 
the circumstellar environment of this star is less dynamic compared to other 
magnetic O-type stars. One of the logical explanations for the detection of 
different magnetic field strengths measured using lines belonging to O, Si, 
and He might be that the lines of these elements are formed at different 
atmospheric heights. We cannot exclude that also Non-Local Thermodynamic 
Equilibrium (NLTE) effects play a role in the line formation of the 
considered elements, causing changes in their equivalent widths. According to 
\citet{zboril},
who estimated the quantitative NLTE effects on the equivalent widths of He 
lines for model atmospheres with an effective temperature of 15000\,K and 
30000\,K, the strength of the \ion{He}{i} lines increases when the NLTE 
effects are taken into account. Studies of NLTE effects were predominantly 
carried out in the past for He-rich magnetic early B-type stars. As an example, 
\citet{gonzalez}
reported that in their study of the strongly magnetic binary HD\,96446 the 
weak \ion{Si}{ii} lines become significantly weaker in NLTE, while \ion{O}{ii}, 
strong \ion{Si}{iii} and weak \ion{Si}{iv} lines become stronger.


\section{Numerical model of the stellar magnetosphere}
\label{sec:MHD}

\begin{figure}
  \begin{center}
    \includegraphics[width=0.4\textwidth]{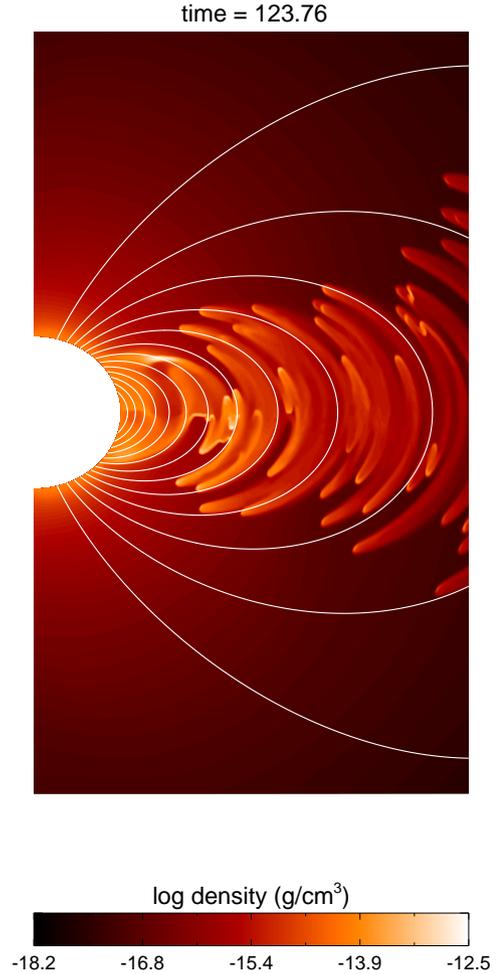}
  \end{center}
  \caption{
Snapshots from a numerical simulation of the wind originating from HD\,54879.
The colour contour plot shows mass density, the white lines the field geometry.
The unit in the time stamp is ks.
  }
  \label{hd54879_sim}
\end{figure}

Current numerical models based on standard magnetohydrodynamics can make 
predictions for the mass distribution around the star, but not for 
concentrations of specific elements. The different field strengths found for 
different elements can therefore not be explained by these models. Assuming 
a dipolar field with a polar field strength of 2\,kG, the star is 
expected to be surrounded by an extensive magnetosphere. This is confirmed 
by numerical simulations. We have rerun the model described in 
\citet{Hubrig2019} 
with a full polar field strength of $B_0= 2$\,kG and 
$\dot{M_0}=10^{-9}\,M_\odot$\,yr$^{-1}$. Figure~\ref{hd54879_sim} shows the 
field structure and density distribution. While far away from the star the 
magnetic field is stretched out by the wind and becomes a split monopole, 
the magnetic field is dominant in the immediate vicinity of the star and the 
dipole structure is preserved. This implies a $1/r^3$ dependence of the 
magnetic field strength on the distance from the stellar surface. The 
smaller field strength found for He places it way above the photosphere.

As Fig.~\ref{hd54879_sim}, shows, the gas flowing from the star escapes 
almost freely at the polar caps, while there is a buildup of gas from 
the magnetic equator to mid-latitudes. However, we observe near the star a 
region of enhanced gas density that fills the volume inside the field lines
up to rather high latitudes. It is thus possible that this region of 
enhanced gas density is the source of the He lines for which we measure 
lower longitudinal field strengths. To make detailed predictions about a 
possible stratification of elements in the magnetosphere, a multi gas 
approach would be needed. This is beyond the scope of the current paper, 
but might be addressed in the future with the latest version of the 
{\sc nirvana} code.

\begin{figure}
  \begin{center}
    \includegraphics[width=0.4\textwidth]{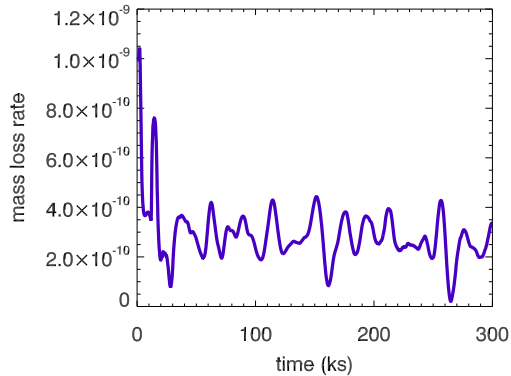}
  \end{center}
  \caption{
Mass loss rate in units of solar masses per year averaged over the 
outer 25~per cent of the simulation box vs.~time.
  }
  \label{massloss}
\end{figure}

Figure~\ref{massloss} shows the mass loss rate as a function of time for the 
model shown in Fig.~\ref{hd54879_sim}. The time series shown in the figure 
has been computed by integrating the mass flux over all latitudes and 
averaging over the outer 25~per cent of the simulation box. After the initial 
phase, the mass loss rate oscillates around a value close to the value of 
$3\times 10^{-10}$ solar masses per year.

\section{Binarity}
\label{sec:bin}

\begin{figure}
 \centering
        \includegraphics[width=\columnwidth]{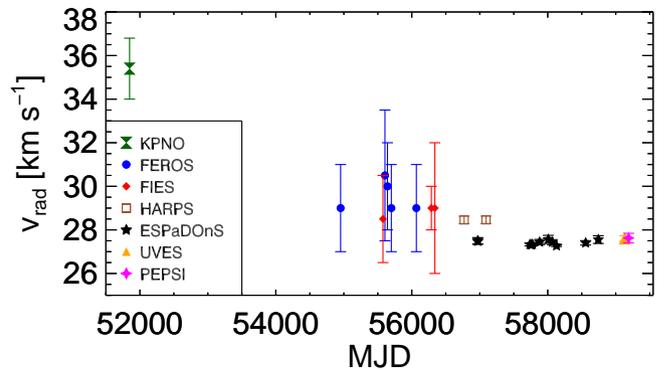}
                \caption{
Radial velocity measurements of HD\,54879 between 2000 and 2020 compiled 
from the literature and supplemented by measurements on the spectra obtained 
with HARPS, ESPaDOnS, UVES, and PEPSI. Our own measurements are carried out 
using LSD Stokes $I$ profiles calculated with the mask containing metal 
lines. }
   \label{fig:rv}
\end{figure}

It is not clear yet whether HD\,54879 is a member of a binary or multiple 
system.
\citet{HD54_2020}
discussed the radial velocity (RV) measurements presented in the literature 
supplemented by own measurements using the LSD Stokes~$I$ spectra calculated 
for HARPS and ESPaDOnS observations. The new RV measurements using the newly 
recorded ESPaDOnS, UVES, and PEPSI spectra indicate a possible increase in 
RV by about 300\,m\,s$^{-1}$ between 2018 January and 2020 December. The 
plot with literature and our own measurements is presented in 
Fig.~\ref{fig:rv}. 
The oldest measurement 
\citep{KPNO}
in this plot is based on observations obtained with the Kitt Peak National 
Observatory (KPNO) Coud\'e feed telescope. The following literature values 
\citep{Castro2015}
are based on data obtained either with the FIbre-fed Echelle Spectrograph 
(FIES) at the Nordic Optical telescope or with the Fiber-fed Extended Range 
Optical Spectrograph (FEROS) at the ESO 2.2\,m telescope.

Both \emph{Hipparcos} and \emph{Gaia}'s second data release
\citep[GDR2;][]{gaia2016,gaia2018} 
data do not raise a binarity flag. On the other hand, HD\,54879 was included 
in the study by 
\citet{Kervella},
who analysed the proper motions of stars presented in the 
\emph{Hipparcos} catalogue
\citep{vanLeeuwen}
and GDR2 data. According to this study, HD\,54879 shows a proper motion 
anomaly, indicating the presence of a companion.

We also made use of archival VLTI observations with PIONIER 
\citep{LeBouquin2011}
to explore whether HD\,54879 has a companion. We reduced the PIONIER data 
with the {\sc pndrs} pipeline
\citep{LeBouquin2011}.
To determine the transfer function applied to calibrate the squared 
visibility amplitudes and the closure phases of the science target, the 
{\sc pndrs} pipeline uses the diameters listed in the JSDC catalog 
\citep{Chelli2016} 
for the interferometric calibrator star, which is observed before and after 
the science target.

We looked for potential companions, utilizing the Python tool
{\sc candid}\footnote{https://github.com/amerand/CANDID}, developed by
\citet{Gallenne2015}.
{\sc candid} performs a search on an adaptive grid with a spacing that will 
allow it to find the global minimum in $\chi^2$. For assessing the 
significance of a companion detection, not only the reduced $\chi^2$ is 
considered, but also the number of degrees of freedom. We scrutinized the 
distance range from 1.6 to 40\,mas around the primary, which, given the 
0.8633\,mas Gaia parallax of HD\,54879, corresponds to projected linear 
separations from 2 to 46\,au. {\sc candid} did not find any companion to 
HD\,54879 up to 3.5 magnitudes fainter than the primary star in the search 
area.

\section{Short-term spectral variability}
\label{sec:pul}

\begin{figure*}
 \centering 
\includegraphics[width=.8\textwidth]{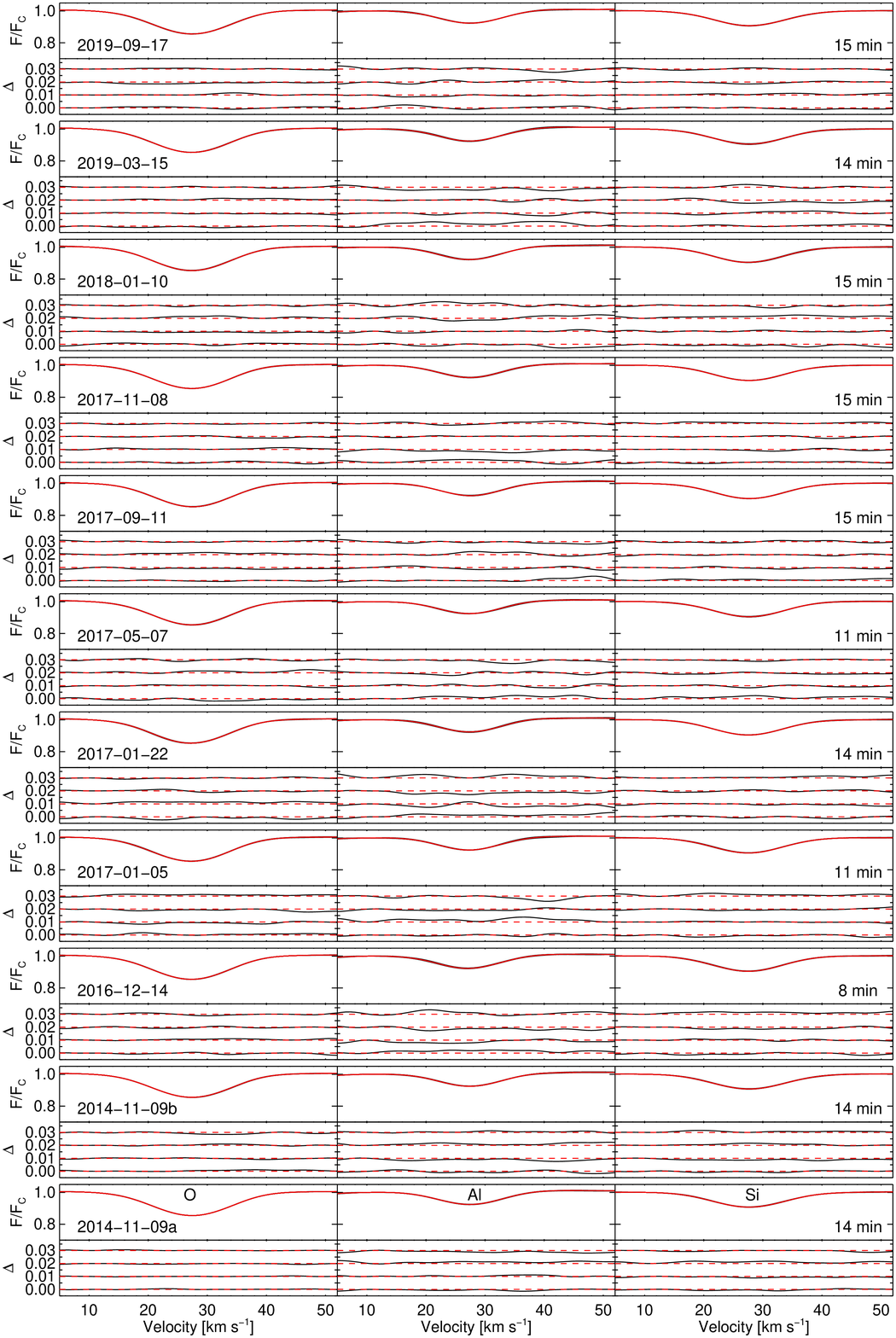}
\caption{ 
LSD Stokes~$I$ profiles calculated for individual ESPaDOnS subexposures 
observed at different epochs. The profiles calculated using oxygen lines are 
shown on the left, those using aluminium and silicon lines are shown in the 
middle and on the right. The upper panels show the overplotted LSD Stokes~$I$ 
profiles calculated for individual subexposures with the mean profiles 
indicated by red lines. In the lower panels we show the differences between 
the LSD Stokes~$I$ profiles for each subexposure and the mean profiles. We 
also indicate the date of the observation and the length of each subexposure.
         }
   \label{fig:Puls}
\end{figure*}

\citet{blomme}
reported that early-type O stars typically show stochastic low-frequency 
variability (SLF), whereas late-type O and early-type B stars typically show 
coherent pulsation modes and suggested that the transition between these 
types of variability takes place at a spectral type around O8. Recently,
\citet{Burssens}
studied the variability of 98 massive O and B stars using the Transiting 
Exoplanet Survey Satellite 
\citep[TESS;][]{TESS}
Sectors 1-13 observations and reported the possible detection of coherent 
$p$ and $g$ modes in a number of main-sequence stars with spectral types in 
the range O4.5-O9.5, including HD\,54879. The authors concluded that the 
dominant periodicity detected in TESS observations does not agree with 
constraints on the rotational period of HD\,54879, hence some other 
mechanism must be responsible.
 
Spectral variability on a short-time scale in HD\,54879 was investigated by
\citet{silva2017}
using subexposures with different integration times in high-resolution HARPS 
spectropolarimetric observations. The analysis of these observations 
indicated distinct changes in line profiles belonging to different elements 
and taking place on timescales corresponding to the duration of the 
individual subexposures of the order of 15--45 min. However, although 
distinct variability was detected, the level of the variability amplitude 
was rather low, between 0.2 and 0.5~per cent. As of today, the only magnetic O-type 
star showing pulsating variability using photometric observations is 
NGC\,1624-2. Using TESS two minute cadence data, 
\citet{kurtz} 
detected three significant low-frequency peaks probably corresponding to 
coherent $g$ mode pulsations. The presence of short-term spectral 
variability in HD\,54879 was also discussed in the study by 
\citet{Hubrig2019},
who reported that several line profiles show distinct radial velocity 
shifts in FORS\,2 spectra. 

Since we have now more extensive spectroscopic material at our disposal,
to test the presence of spectral variability on short time scales of the 
order of minutes, we used all available high-resolution ESPaDOnS 
spectropolarimetric observations, consisting each of four subexposures 
recorded with integration times between 8 and 15\,min, which are even 
shorter than the subexposure integration times with HARPS. As 
\citet{silva2017} 
reported the strongest variations -- of up to 0.5~per cent -- in line profiles 
calculated from HARPS spectra using exclusively silicon lines, we compared 
the variability amplitude of the silicon lines with the variability of line 
profiles belonging to the elements oxygen and aluminium. The line mask for 
oxygen used in this test included fourteen \ion{O}{ii} lines, that for Al 
four \ion{Al}{iii} lines, and the silicon mask included five \ion{Si}{iii} 
and two \ion{Si}{iv} lines. The LSD profiles calculated for these masks for 
all individual ESPaDOnS subexposures are presented in Fig.~\ref{fig:Puls}. 
The lowest amplitude of variability is detected in oxygen lines with an 
intensity amplitude in the range between 0.07 and 0.24~per cent. The least 
variable LSD line profiles belong to the first data set from 2014 November~9, 
whereas the most variable profiles are observed in the data set from 2017 
January~5. The aluminium and silicon lines show stronger variability with 
amplitudes 0.21--0.36~per cent and 0.15--0.26~per cent, respectively. The 
lowest amplitude for aluminium was again detected in the first data set 
obtained on 2014 November~9, whereas the LSD silicon profiles were least 
variable in 2017 September~11. The strongest variability amplitude is 
observed in the Al lines in the data set from 2019 March~3 and that for 
the Si lines in the data set from 2018 January~10. To conclude, the 
amplitude of the line profile variability is slightly different for 
different elements, but is always below 0.5~per cent. 


\section{Discussion} 
\label{sec:dis}

In agreement with previous studies of HD\,54879 
\citep{Shenar2017,Hubrig2019,HD54_2020},
our recently acquired spectropolarimetric observations show a very slow 
magnetic field variability related to the extremely slow rotation of 
HD\,54879. The extremely slow rotation of this star is also indicated in our 
dynamical spectrum, displaying the variability of the H$\alpha$ line. From 
the end of 2017 to the end of 2019, we were observing the star at rotational 
phases with the best visibility of the magnetic equator, leading to a low 
positive \bz{} value around 50\,G in 2019 March. At the end of 2020, the 
field strength decreased to about $-$300\,G, measured in the LSD Stokes~$V$ 
spectra calculated for metal lines. The strongest field in HD\,54879 of the 
order of $-$570\,G was observed in 2014, but it is not clear yet whether 
this field strength indeed represents the extremum of the field of negative 
polarity or if the field will reach an even higher strength when the full 
rotation cycle is covered by future spectropolarimetric observations. 

The question on the presence of a companion to HD\,54879 is still an open 
important issue in view of the scenario that magnetic fields in massive stars 
may be generated by strong  binary interaction and a recent report by 
\citet{frost}
that the primary component of the magnetic Of?p star HD\,148937 is most 
likely a rejuvenated merger product. Our radial velocity study indicates 
that if HD\,54879 is indeed a binary member, the orbital period should be 
very long. As of today, only the study of 
\citet{Kervella}, 
who detected a proper motion anomaly, indicates the possible presence of a 
companion.

The most intriguing result of our study is the discovery of differences in 
longitudinal  magnetic field strengths measured using different LSD masks 
containing  lines belonging to different elements. It is the first time that 
such a differential analysis of the field strengths in dependence of the 
used lines is carried out for a magnetic O-type star. While in magnetic Ap 
and Bp stars chemical elements have different distributions across the 
stellar surface and sample the magnetic field topology in different manners, 
the presence of chemical spots is not expected in O-type stars with 
radiatively driven winds. Interestingly, conspicuously differing field 
strengths measured using individual lines belonging to different elements 
were reported for the strongly magnetic Of?p star NGC\,1624-2 by 
\citet{Wade2012}.
Although the authors noticed that their measurements exhibit a scatter 
larger than predicted by the standard deviations, they attempted to explain 
the differences in the measurements by difficulties of establishing the 
position of the centre of gravity in the relatively noisy spectra. 

In our study of HD\,54879 remarkable differences in the measurements using 
line masks for the elements O, Si, and He are detected on a few epochs close 
to the best visibility of the negative magnetic field pole: the field 
strengths using the He lines systematically lower than the field strengths 
measured using other masks and the field strengths using the O lines are 
systematically higher than the field strengths measured using the Si lines. 
In contrast to the usually observed variability of the Stokes~$I$ profiles 
in Ap and Bp stars showing surface patchiness, the LSD Stokes~$I$ profiles 
of the studied line masks remain stable over all observed epochs from 2014 
to 2020, and are not noteworthy contaminated by emission from the stellar 
wind or material trapped in the magnetosphere. Thus, we conclude that the 
elements O, Si, and He are not uniformly distributed over the stellar 
surface of HD\,54879, but that the formation of their spectral lines takes 
place at different atmospheric heights, with He lines formed much higher in 
the stellar atmosphere compared to the Si and O lines. Interestingly,
\citet{Castro2015}
considered the abundances of carbon, nitrogen, oxygen, silicon, and
magnesium of HD\,54879 and reported that they are slightly lower than solar,
but comparable within the errors.
\citet{Martins2015}
studied the carbon, nitrogen, and oxygen abundances in 74 massive O-type
stars, also including seven O stars with known  magnetic fields in the 
sample, and concluded that their surface  abundances do not deviate from
other non-magnetic O-type stars. They did not take into account a potential
vertical abundance stratification.

Obviously, the detected element stratification has to be considered in the 
context of the dynamic competition between the outward forces of radiation 
pressure and the channelling and confinement of an outflowing wind by the 
large-scale organized magnetic field. Future magnetohydrodynamical 
simulations have to include chemical gradients in the wind plasma 
to explain the detected vertical abundance stratification of different 
elements in hot stars possessing dynamical magnetospheres. We also discuss 
the possible contribution of NLTE effects, causing changes in the equivalent 
widths of lines belonging to different elements. Obviously, NLTE modeling 
should be implemented in the line forming calculations in studies of 
vertical element stratification in O-type stars.

Short-term spectral variability is certainly present in the high-resolution
spectropolarimetric observations of HD\,54879, but the variability amplitude
in the normalised intensity is rather low, not higher than 0.5~per cent. 
Pulsations in massive stars are commonly associated with coherent pulsation 
modes triggered by an opacity mechanism operating in the Z-bump, relating to
iron-peak elements 
\citep[e.g.][]{dziembowski, miglio}.
Asteroseismological studies allowing to directly access stellar interiors
gained momentum in recent years due to the availability of photometric TESS
data. The vertical abundance stratification of different elements in magnetized
atmospheres is a known phenomenon and was well-documented for Bp stars and,
in particular, for rapidly oscillating Ap (roAp) stars
\citep[e.g.][]{Hubrig2007,Cowley2001}.
While the reason for stratification in roAp  stars is related to the
existence of a non-standard temperature gradient in the pulsating atmospheric
layers, the origin of the element stratification in magnetic Bp stars and 
HD\,54879 is not clear. On the other hand, since 
\citet{Burssens} 
reported on the detection of low-frequency variability in the TESS 
photometric data of several magnetic massive stars including HD\,54879, it 
is possible that pulsations can affect the chemical structure also in 
atmospheres and magnetospheres of magnetic O-type stars.

There exist some pieces of evidence that pulsations play an important role in 
the structure of the emitting plasma close to the photosphere in a massive 
star. A couple of years ago, 
\citet{Oskinova}
reported that at least one $\beta$~Cephei-variable, the extremely slowly 
rotating magnetic B0.7\,IV star $\xi^{1}$\,CMa, shows periodic variability in 
X-rays in phase with the optical pulsational light curve. The mechanism by 
which the X-rays are affected by the stellar pulsation remains 
enigmatic. According to 
\citet{Shenar2017},
the X-ray flux of HD\,54879 recorded with the XMM-Newton satellite indicates 
a higher X-ray luminosity compared to other stars with similar spectral types.
However, the data set that was considered in this work did not cover 
significantly long time intervals to search for short-term variability of 
the X-ray light curve. Similar to HD\,54879, $\xi^{1}$\,CMa hosts a dynamical 
magnetosphere. 
\citet{Shultz}
showed that H$\alpha$ emission in this star is modulated by both pulsation 
and rotation. Although short- and long-term variability is detectable in the 
H$\alpha$ line in the acquired spectra of HD\,54879, our data are not 
sufficient to estimate short-term periodicity. Clearly, long-cadence 
spectroscopic and photometric observations are necessary to get more insight 
into the pulsational variability of HD\,54879.


\section*{Acknowledgements}
We thank the anonymous referee for their comments.
SPJ is supported by the German Leibniz-Gemeinschaft, project number P67-2018.
Based on observations made with ESO Telescopes at the La Silla Paranal 
Observatory under programme IDs~0102.D-0234(C), 0106.D-0250(A) and 
0106.D-0250(B).
Based on data acquired with the Potsdam Echelle Polarimetric and 
Spectroscopic Instrument (PEPSI) using the Large Binocular Telescope (LBT) 
in Arizona. The LBT is an international collaboration among institutions in 
the United States, Italy, and Germany. LBT Corporation partners are the 
University of Arizona on behalf of the Arizona university system; Istituto 
Nazionale di Astrofisica, Italy; LBT Beteiligungsgesellschaft, Germany, 
representing the Max-Planck Society, the Leibniz-Institute for Astrophysics 
Potsdam (AIP), and Heidelberg University; the Ohio State University; and 
the Research Corporation, on behalf of the University of Notre Dame, 
University of Minnesota, and University of Virginia.
Based on observations collected at the Canada-France-Hawaii Telescope (CFHT),
which is operated by the National Research Council of Canada, the Institut
National des Sciences de l'Univers of the Centre National de la Recherche
Scientifique of France, and the University of Hawaii.


\section*{Data Availability}

The data obtained with ESO facilities will be available in the ESO Archive at
http://archive.eso.org/ and can be found with the instrument and
object name.
The ESPaDOnS data underlying this article are available in the CFHT Science 
Archive at https://www.cadc-ccda.hia-iha.nrc-cnrc.gc.ca/en/cfht/ and can be 
accessed with the instrument and object name.
The PEPSI data underlying this article will be shared on a reasonable
request to the corresponding author.


\appendix
\section{Profiles for longitudinal magnetic field measurements}

\begin{figure*}
 \centering 
\includegraphics[width=0.66\textwidth]{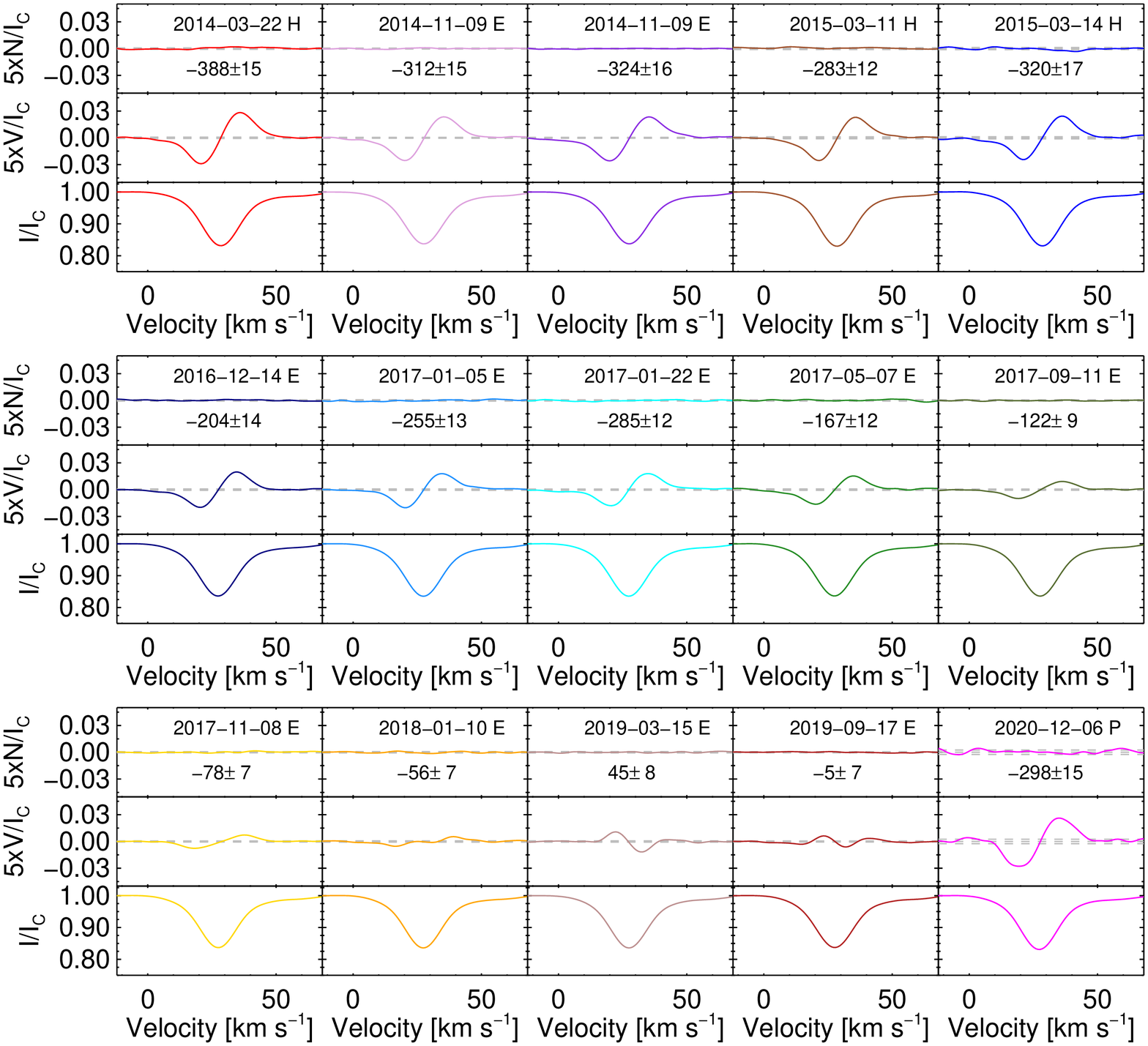}
\caption{
LSD Stokes $I$, $V$, and null profiles for the line list having all lines. 
The plots are sorted by observing date and the instrument used is indicated 
after the date as follows: H=HARPS, E=ESPaDOnS, and P=PEPSI. 
         }
   \label{afig:LSDall}
\end{figure*}

\begin{figure*}
 \centering 
\includegraphics[width=0.66\textwidth]{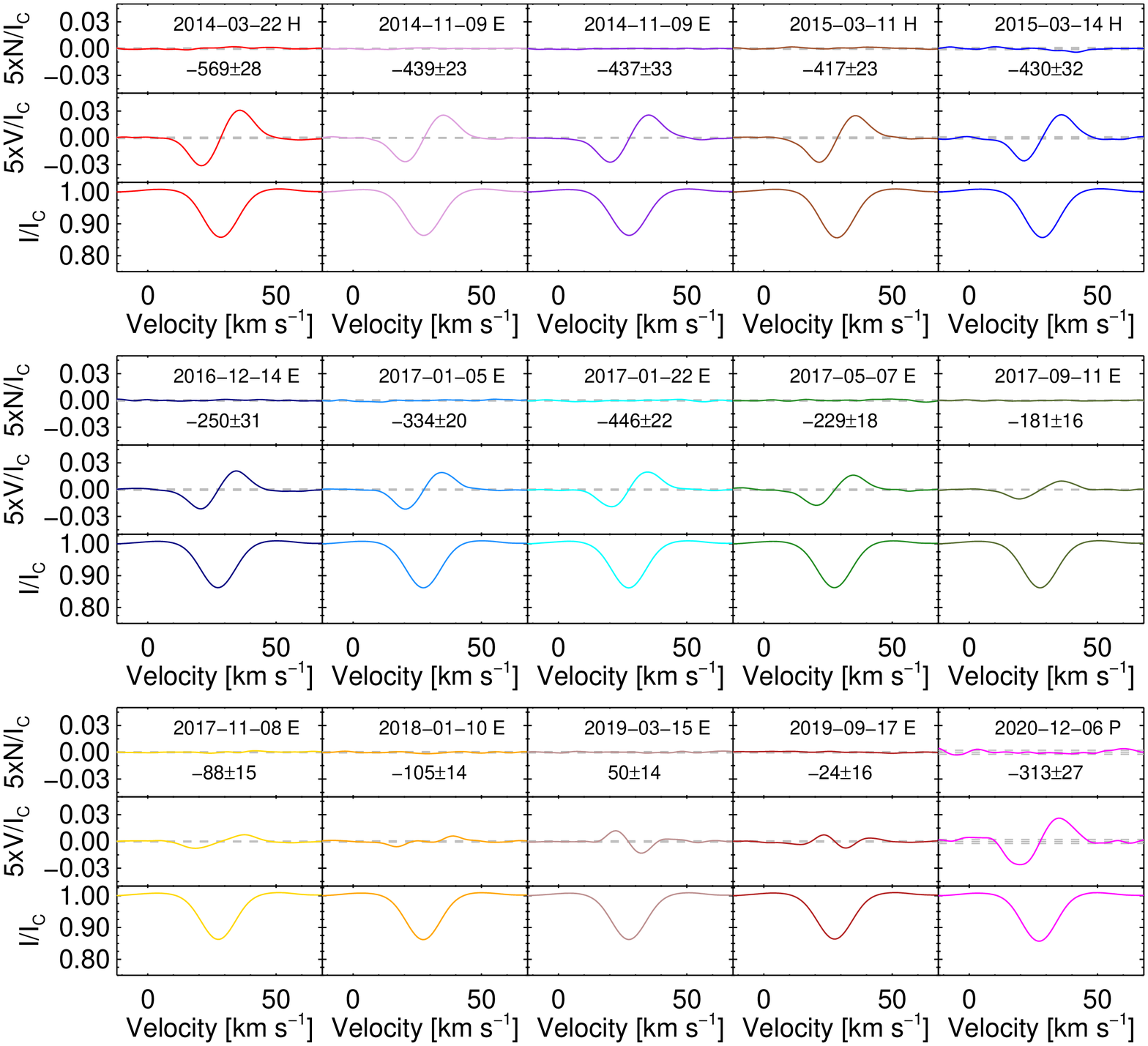}
\caption{As Fig.~\ref{afig:LSDall}, but using only metal lines.
         }
   \label{afig:LSDmet}
\end{figure*}

\begin{figure*}
 \centering 
\includegraphics[width=0.66\textwidth]{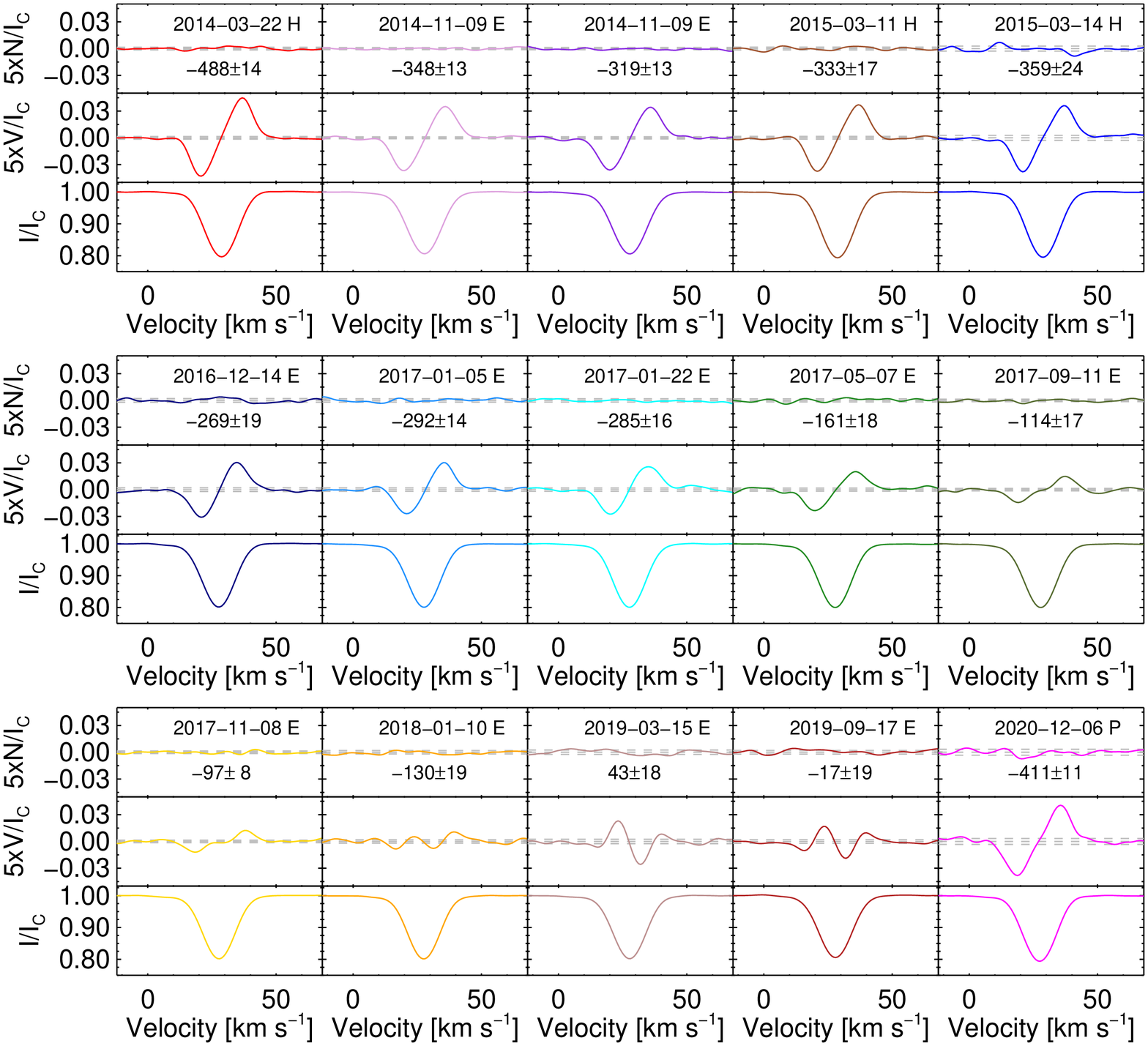}
\caption{As Fig.~\ref{afig:LSDall}, but using only silicon lines.
         }
   \label{afig:LSDSi}
\end{figure*}

\begin{figure*}
 \centering 
\includegraphics[width=0.66\textwidth]{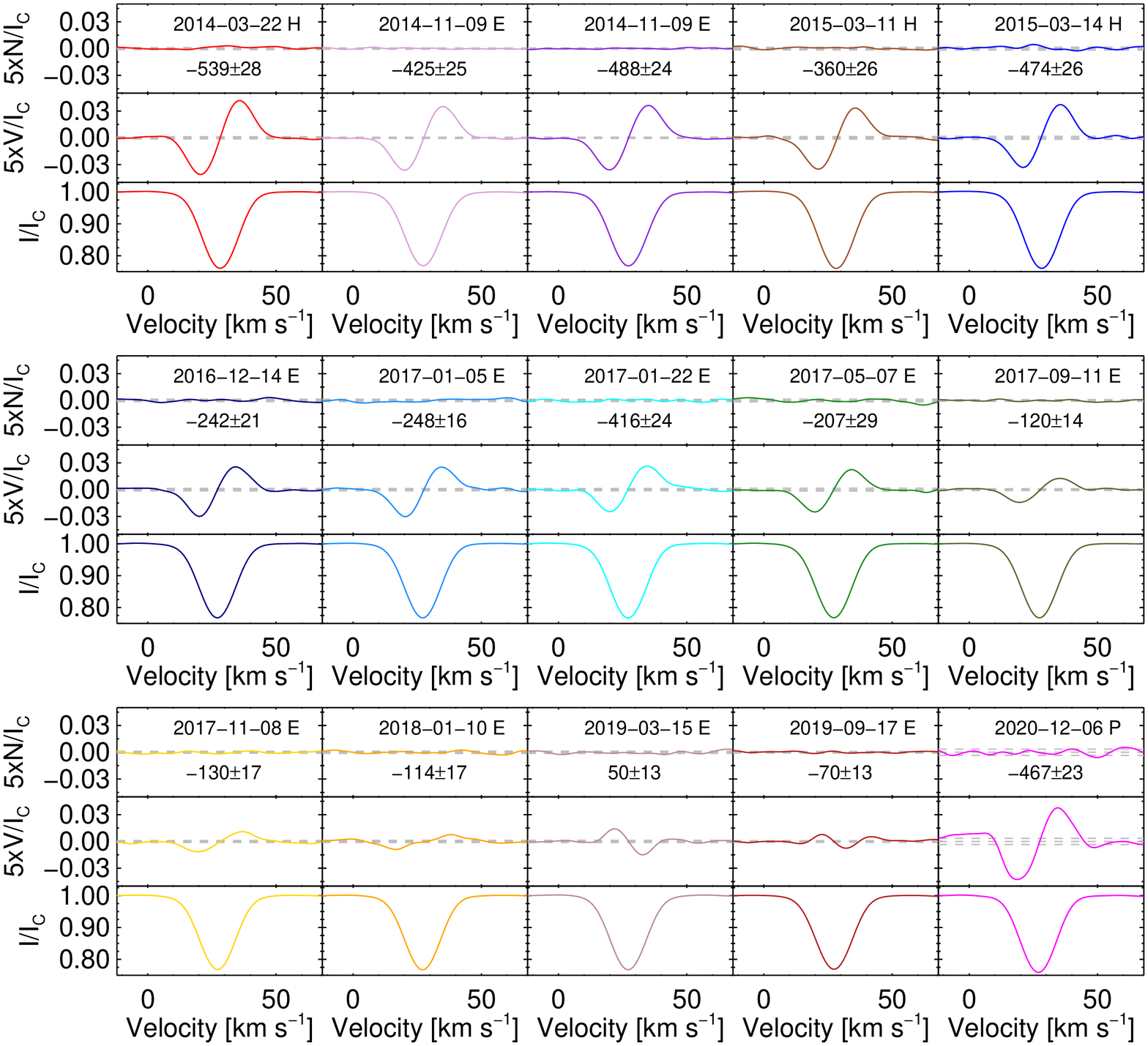}
\caption{As Fig.~\ref{afig:LSDall}, but using only oxygen lines.
         }
   \label{afig:LSDO}
\end{figure*}

\begin{figure*}
 \centering 
\includegraphics[width=0.66\textwidth]{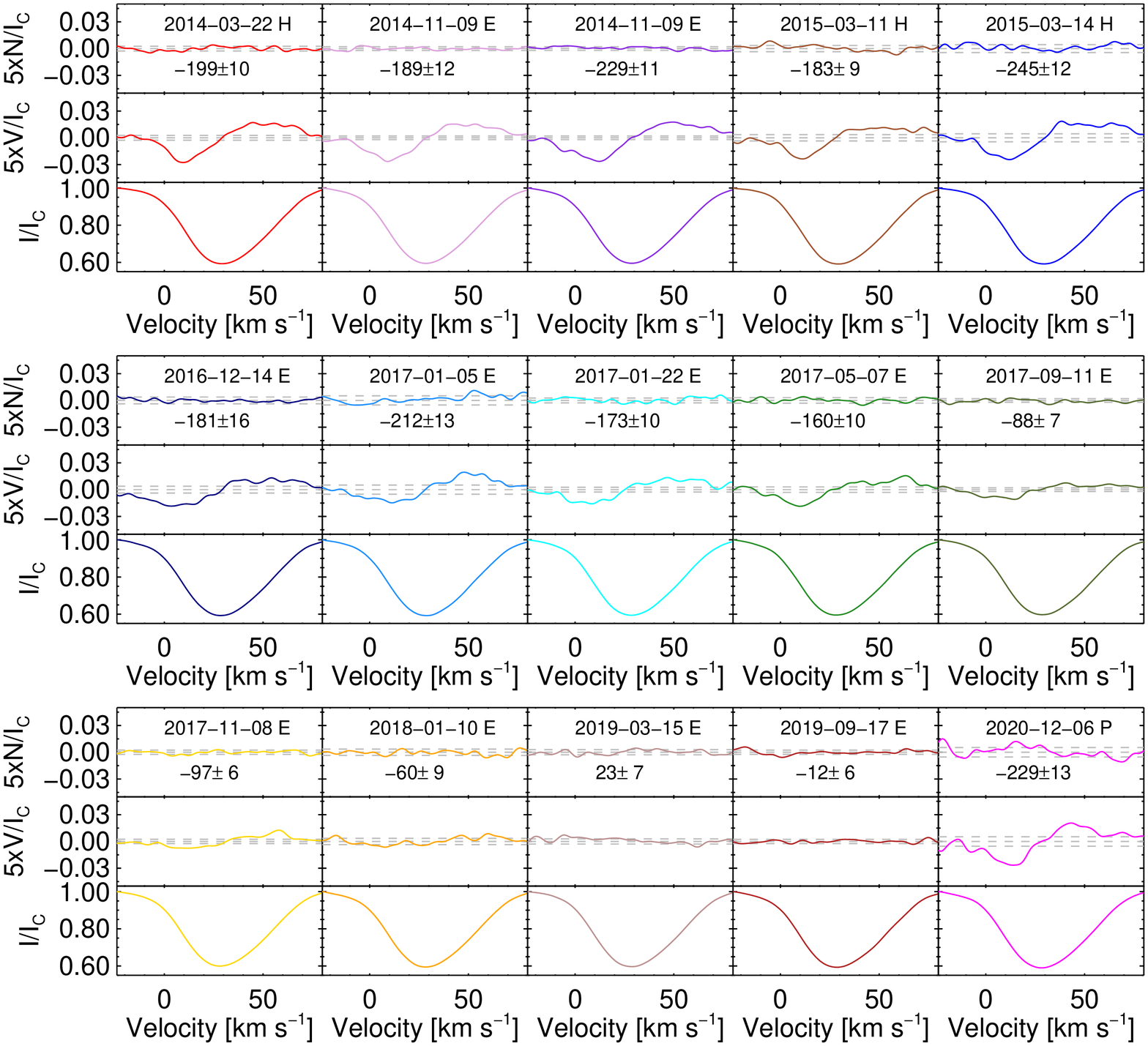}
\caption{As Fig.~\ref{afig:LSDall}, but using only neutral helium lines.
         }
   \label{afig:LSDHe1}
\end{figure*}

LSD Stokes~$I$, $V$, and null profiles are presented
in Fig.~\ref{afig:LSDall} for the line list having both metal and helium 
lines,
in Fig.~\ref{afig:LSDmet} for the line list having only metal lines, 
in Fig.~\ref{afig:LSDSi} for the line list having only twice ionized silicon 
lines, 
in Fig.~\ref{afig:LSDO} for the line list having only once ionized oxygen 
lines, 
and in Fig.~\ref{afig:LSDHe1} for the line list having only neutral hydrogen 
lines.
The plots are sorted by observing epoch.
Shaded regions in Stokes~$V$ and null panels indicate the mean uncertainty.

\label{lastpage}

\end{document}